\documentclass[journal]{IEEEtran}
\usepackage{graphics} 
\usepackage{epsfig} 
\usepackage{times} 
\usepackage{amsmath} 
\usepackage{amssymb}  
\usepackage{cite}
\usepackage{epstopdf}
\usepackage{subfigure,color,balance,framed}
\usepackage{verbatim}
\usepackage{cases}
\usepackage{enumerate}
\usepackage{booktabs}
\usepackage{balance}
\usepackage{mathbbol}
\usepackage{algorithm}
\usepackage{algorithmicx}
\usepackage{algpseudocode}
\usepackage{mathrsfs}
\usepackage{threeparttable}

\newcommand\markend{{\hfill {\small$\square$}}}

\newcommand{\method}[1]{\texttt{#1}}

\usepackage[colorlinks=true,      
linkcolor=black,      
citecolor=black,      
filecolor=black,      
urlcolor=blue]{hyperref}

\newtheorem{definition}{Definition}
\newtheorem{theorem}{Theorem}
\newtheorem{proposition}{Proposition}
\newtheorem{remark}{Remark}
\newtheorem{lemma}{Lemma}
\newtheorem{corollary}{Corollary}

\newtheorem{assumption}{Assumption}

\def\Y{{\bf Y}}

\def\0{{\bf 0}}
\def\1{{\bf 1}}

\def\col{{\mathrm {col}}}
\def\st{{\mathrm {subject~to}}}

\def\eg{{\em e.g.}}
\def\ie{{\em i.e.}}

\def\diag{{\rm diag}}

\begin{document}
%
\title{
{DeeP-LCC}: Data-EnablEd Predictive Leading Cruise Control in Mixed Traffic Flow
}

\author{Jiawei Wang, Yang Zheng, Keqiang Li and Qing Xu
	\thanks{The work of J. Wang, K. Li, and Q. Xu is supported by National Key R\&D Program of China with 2018YFE0204302,  National Natural Science Foundation of China with 52072212, and Tsinghua University-Didi Joint Research Center for Future Mobility. All correspondence should be sent to Y.~Zheng and Q.~Xu.}
	\thanks{J.~Wang, K.~Li and Q.~Xu are with the School of Vehicle and Mobility, Tsinghua University, Beijing 100084, China. (wang-jw18@mails.tsinghua.edu.cn, \{likq,qingxu\}@tsinghua.edu.cn).}%
	\thanks{Y. Zheng is with the Department of Electrical and Computer Engineering, University of California San Diego, CA 92093, USA. ({zhengy@eng.ucsd.edu}).}%
}

\maketitle

\begin{abstract}
	
For the control of connected and autonomous vehicles (CAVs), most existing methods focus on model-based strategies. They require explicit knowledge of car-following dynamics of human-driven vehicles that are non-trivial to identify accurately. In this paper, instead of relying on a parametric car-following model, we introduce a data-driven non-parametric
strategy, called \method{DeeP-LCC} (Data-EnablEd Predictive Leading Cruise Control), to achieve safe and optimal control of CAVs~in mixed traffic. We first utilize Willems' fundamental lemma~to~obtain a data-centric  representation of mixed traffic behavior. This is justified by rigorous analysis on controllability and observability properties of mixed traffic. We then employ a receding horizon strategy to solve a finite-horizon optimal control problem at each time step, in which input/output constraints are incorporated for collision-free guarantees. Numerical experiments validate~the performance of \method{DeeP-LCC} compared to a standard predictive controller that requires an accurate model. { Multiple nonlinear traffic simulations further confirm its great potential on improving traffic efficiency, driving safety, and fuel economy. }
	
\end{abstract}


\section{Introduction}

\IEEEPARstart{W}{ireless} communication technologies, \eg, vehicle-to-vehicle (V2V) or vehicle-to-infrastructure (V2I), have provided new opportunities for advanced vehicle control and enhanced traffic mobility~\cite{karagiannis2011vehicular}. With access to beyond-the-sight information and edge/cloud computing resources,~individual vehicles are capable to make sophisticated decisions and even cooperate with each other to achieve system-wide traffic optimization. One typical technology is Cooperative Adaptive Cruise Control (CACC), which organizes a series of connected and autonomous vehicles (CAVs) into a platoon and applies cooperative control strategies to achieve smaller spacing, better fuel economy,~and~smoother~traffic flow~\cite{li2017dynamical,zheng2016stability,milanes2013cooperative}.

In practice, CACC or platooning requires all the involved vehicles to have autonomous capabilities. Considering the gradual deployment of CAVs, the transition phase of mixed traffic with the coexistence of human-driven vehicles (HDVs) and CAVs may last for decades~\cite{stern2018dissipation,zheng2020smoothing,li2022cooperative}. HDVs, connected to V2V/V2I communication but still controlled by human drivers, will still be the majority on public roads in the near future. Without explicitly considering surrounding HDVs' behavior,  CAVs at a low penetration rate may only bring negligible benefits on traffic performance~\cite{shladover2012impacts,talebpour2016influence}. 
One extension of CACC to mixed traffic is Connected Cruise Control (CCC)~\cite{orosz2016connected}, in which one single CAV at the tail makes its control decisions by exploiting the information of multiple HDVs ahead. Another recent extension is Leading Cruise Control (LCC) that incorporates the motion of HDVs ahead and behind~\cite{wang2021leading}. 


Existing CAV control, \eg, CACC and CCC, mainly takes local-level performance into consideration --- the CAVs aim to improve their own driving performance. Considering the interactions among surrounding vehicles, a recent concept of Lagrangian control in mixed  traffic aims to focus on system-level performance of the entire traffic flow by utilizing CAVs as \textit{mobile actuators}~\cite{stern2018dissipation,zheng2020smoothing,vinitsky2018lagrangian}. In particular, the real-world experiment in \cite{stern2018dissipation} demonstrates the potential of one single CAV in stabilizing a ring-road mixed traffic system. This has been subsequently validated from rigorous theoretical analysis~\cite{zheng2020smoothing,wang2020controllability} and large-scale traffic simulations~\cite{wu2021flow,vinitsky2018lagrangian}. These results focus on a closed circular road setup~\cite{sugiyama2008traffic}. The recent notion of LCC~\cite{wang2021leading} focuses on general open straight road scenarios and has provided further insight into CAV control in mixed traffic: one single CAV can not only adapt to the downstream traffic flow consisting of its preceding HDVs (as \textit{a follower}), but also improve the upstream traffic performance by actively leading the motion of its following HDVs (as \textit{a leader}). This explicit consideration of a CAV as both a leader and a follower greatly enhances its capability in smoothing mixed traffic flow, as demonstrated both empirically and theoretically in~\cite{wang2021leading}. One challenge is to design LCC strategies  with safety guarantees in smoothing traffic flow when the traffic model is not known. 


\subsection{Model-Based and Model-Free Control of CAVs}

{ Mixed traffic is a complex human-in-the-loop cyber-physical system, in which HDVs are controlled by human drivers with uncertain and stochastic behaviors.   
Most~existing studies exploit microscopic car-following models 
and design model-based control strategies for CAVs, such as linear quadratic control~\cite{jin2017optimal,zheng2020smoothing}, structured optimal control~\cite{wang2020controllability}, $\mathcal{H}_\infty$ control~\cite{zhou2020stabilizing} and model predictive control~\cite{zheng2016distributed}. 
In practice, however, human car-following behaviors are complex and nonlinear, which are non-trivial to identify accurately.} Model-free and data-driven methods, 
bypassing model identifications, have recently received increasing attention~\cite{recht2019tour,furieri2020learning}. For example, reinforcement learning~\cite{vinitsky2018lagrangian,wu2021flow} and adaptive dynamic programming~\cite{gao2016data,huang2020learning} have been recently utilized for mixed traffic control. Instead of relying on explicit dynamics of HDVs, these methods utilize  online and/or offline driving data of HDVs to learn  CAVs' control strategies. However, these methods typically bring a heavy computation burden and are sample inefficient. Safety is a critical aspect for CAV control in practical deployment, but this has not been well 
addressed in the existing studies~\cite{vinitsky2018lagrangian,wu2021flow,gao2016data,huang2020learning}. Indeed, it remains challenging to include constraints to achieve safety guarantees for these model-free and data-driven methods~\cite{recht2019tour}.

On the other hand, model predictive control (MPC) has been widely recognized as a primary tool to address control problems with constraints~\cite{camacho2013model,zheng2016distributed}. Recent advancements in data-driven MPC have further provided techniques towards safe learning-based control using measurable data~\cite{lan2021data,hewing2020learning,coulson2019data}. 
One promising method is the Data EnablEd Predictive Control (DeePC)~\cite{coulson2019data} that is able to achieve safe and optimal control for unknown systems using input/output measurements. Rather than identifying a parametric system model, DeePC relies on Willems' \emph{fundamental lemma}~\cite{willems2005note} to directly learn the system behavior and predict future trajectories. In particular,  DeePC 
allows one to incorporate input/output constraints to ensure safety. 
It has been shown theoretically that DeePC is equivalent to sequential system identification and MPC for deterministic linear time-invariant (LTI) systems~\cite{coulson2019data,fiedler2021relationship}, and empirically that DeePC could achieve comparable control performance with respect to MPC with accurate model knowledge for stochastic and nonlinear systems~\cite{coulson2019regularized,dorfler2022bridging}. Recently, practical applications have been seen in quadcopter systems~\cite{elokda2021data}, power~grids~\cite{huang2021decentralized}, and electric motor drives~\cite{carlet2020data}. 
%

To our best knowledge, data-driven MPC methods, particularly the recent  DeePC method, have not been discussed for mixed traffic control. Due to distinct and complex dynamical properties of mixed traffic systems, the aforementioned results~\cite{coulson2019data,carlet2020data,huang2021decentralized} are not directly applicable.

\subsection{Contributions}

In this paper, we focus on the recent LCC framework~\cite{wang2021leading} and design safe and optimal control strategies for CAVs to smooth mixed traffic flow. Our method requires no prior knowledge of HDVs' car-following dynamics.~In particular, we introduce a Data-EnablEd Predictive Leading Cruise Control (\method{DeeP-LCC}) strategy, in which the CAVs utilize measurable driving data for controller design with collision-free guarantees. Some preliminary results were presented in~\cite{wang2022data}. Our contributions of this work are as follows.   

We first establish a linearized state-space model for a general mixed traffic system with multiple CAVs and HDVs under the LCC framework. We directly use measurable driving data as system output since the HDVs' equilibrium spacing is typically not measurable. This issue of unknown equilibrium spacing has been neglected in many recent studies on mixed traffic~\cite{jin2017optimal,wang2020controllability,di2019cooperative,gao2016data,huang2020learning,lan2021data}. We further show that the linearized mixed traffic system is not controllable (except the case when the first vehicle is a CAV), but is stabilizable and observable. These results are the foundations of our adaptation~of~DeePC~\cite{coulson2019data} for mixed traffic control.

 We then propose a \method{DeeP-LCC} method for CAV control, which directly utilizes  HDVs' trajectory data and bypasses an explicit identification of a parametric car-following model. The standard  DeePC requires the underlying system to be controllable~\cite{coulson2019data,willems2005note}, and thus cannot be directly applied to mixed traffic. To resolve this, { we introduce an external input signal to record the data of the head vehicle, \ie, the first vehicle at the beginning of the mixed traffic system.} Together with CAVs' control input, this contributes to system controllability. 
	Our \method{DeeP-LCC} formulation incorporates spacing constraints on the driving behavior and thus  provides safety guarantees for CAVs when feasible. 
	Furthermore, our \method{DeeP-LCC} is directly applicable to nonlinear and non-deterministic traffic systems.

 We finally carry out multiple traffic simulations~to~validate the performance of \method{DeeP-LCC}. \method{DeeP-LCC} achieves comparable performance in nonlinear and non-deterministic cases with respect to a standard MPC based on an accurate linearized model. We also design an urban/highway driving scenario motivated by the New European Driving Cycle (NEDC) and an emergence braking scenario. Numerical results confirm the benefits of  \method{DeeP-LCC} in improving driving safety, fuel economy and traffic smoothness. Particularly, \method{DeeP-LCC} reduces up to~$24.69\%$ fuel consumption  with safety guarantees in the braking scenario at a CAV penetration rate of $25\%$ compared with the case of all HDVs. 

\subsection{Paper Organization and Notation}

The rest of this paper is organized as follows. Section~\ref{Sec:2} introduces the modeling for the mixed traffic system, and Section~\ref{Sec:3} presents controllability and observability analysis. This is followed by a brief review of the standard  DeePC in Section~\ref{Sec:4}. We present \method{DeeP-LCC} in Section~\ref{Sec:5}. Traffic simulations are discussed in Section~\ref{Sec:6}. Section~\ref{Sec:7} concludes this paper. Some auxiliary proofs and implementation details are included in the appendix.

\begin{figure*}[t]
	\vspace{1mm}
	\centering
	\hspace{1.5mm}
	\includegraphics[scale=0.45]{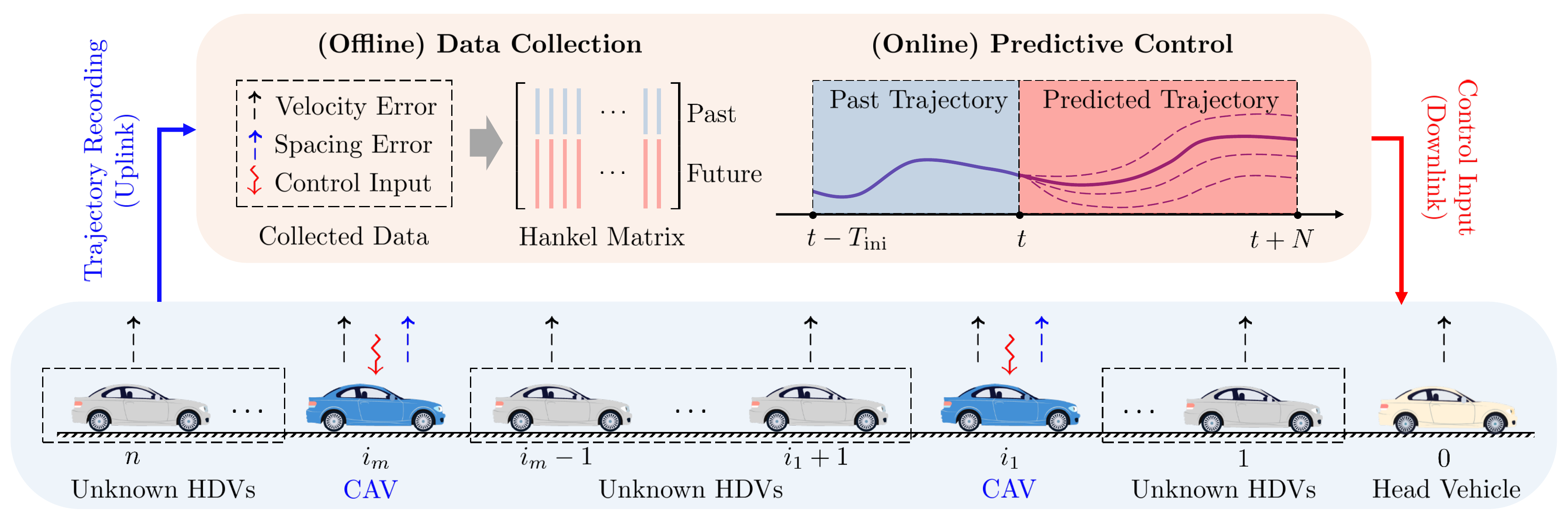}
	\vspace{-1mm}
	\caption{Schematic of \method{DeeP-LCC} for CAVs in mixed traffic. The head vehicle is located at the beginning, indexed as 0, behind which there exist $n$ vehicles indexed from 1 to $n$. The $n$ vehicles consist of $n-m$ HDVs, whose car-following dynamics are unknown, and $m$ CAVs, indexed from $i_1$ to $i_m$. In (offline) data-collection, \method{DeeP-LCC} records measurable data of the mixed traffic system, including velocity errors of each vehicle (represented by the black dashed arrow) and spacing errors of the CAVs (represented by the blue dashed arrow). Then,  \method{DeeP-LCC} utilizes these data to construct Hankel matrices for future trajectory predictions. In (on-line) predictive control, \method{DeeP-LCC} employs the collected data to design the optimal future trajectory and sends the control signal to the CAVs (represented by the red squiggle arrow). Details on \method{DeeP-LCC} are presented in Section~\ref{Sec:5}.}
	\label{Fig:SystemSchematic}
	\vspace{-2mm}
\end{figure*}

\emph{Notations:}
We denote $\mathbb{N}$ as the set of natural numbers, $\mathbb{0}_n$ as a zero vector of size $n$, and $\mathbb{0}_{m\times n}$ as~a~zero matrix of  size $m \times n$. For a vector $a$ and a positive definite matrix $X$, $\|a\|_{X}^{2}$ denotes the quadratic form $a^\top X a$. Given a collection of vectors $a_1,a_2,\ldots,a_m$, we denote $\col(a_1,a_2,\ldots,a_m)=\begin{bmatrix}
a_1^\top,a_2^\top,\ldots,a_m^\top
\end{bmatrix}^\top$. Given matrices of the same column size $A_1,A_2,\ldots,A_m$, we denote $\col(A_1,A_2,\ldots,A_m)=\begin{bmatrix}
A_1^\top,A_2^\top,\ldots,A_m^\top
\end{bmatrix}^\top$. Denote $\diag(x_1,\ldots,x_m)$ as a diagonal matrix with $x_1,\ldots,x_m$ on its diagonal entries, and $\diag(D_1,\ldots,D_m)$ as a block-diagonal matrix with matrices $D_1,\ldots,D_m$ on its diagonal blocks. We use $\mathbb{e}_{n}^i$ to denote a $n \times 1$ unit vector, with the $i$-th entry being one and the others being zeros.  Finally, $A\otimes B$ represents the Kronecker product between matrices $A$ and $B$.

\section{Theoretical Modeling Framework}
\label{Sec:2}

In this section, we first introduce the nonlinear modeling of HDVs' car-following behavior, and then present the linearized dynamics of a general mixed traffic system under the LCC framework~\cite{wang2021leading}. 

{ As shown in Fig.~\ref{Fig:SystemSchematic}, we consider a general mixed traffic system with $n+1$ individual vehicles, among which there exist one head vehicle, indexed as $0$, and $m$ CAVs and $n-m$ HDVs in the following $n$ vehicles, indexed from $1$ to $n$. Note that essentially any HDV ahead of the first CAV can be designated as the head vehicle.} Define $\Omega=\{1,2,\ldots,n\}$ as the index set of all the following vehicles, ordered from front to end, and $S=\{i_1,i_2,\ldots,i_m\}\subseteq \Omega$ as the set of the CAV indices, where $i_1 < i_2 < \ldots < i_m$ also represent the spatial locations of the CAVs in the mixed traffic. The position, velocity and acceleration of the $i$-th vehicle at time $t$ is denoted as $p_i(t)$, $v_i(t)$ and $a_i(t)$, respectively. 

\subsection{Nonlinear Car-Following Dynamics of HDVs}

\label{Sec:NonlinearTraffic}

There are many well-established models to describe  car-following dynamics of HDVs, such as the optimal velocity model (OVM)~\cite{bando1995dynamical}, and the intelligent driver model (IDM)~\cite{treiber2000congested}. These models can capture various human driving behaviors and reproduce typical traffic phenomena, \eg, stop-and-go traffic waves~\cite{treiber2013traffic}. 

In these models, the acceleration of an HDV depends on its car-following spacing $s_i(t)=p_{i-1}(t)-p_i(t)$, \ie, the bumper-to-bumper distance between vehicle $i$ and its preceding vehicle $i-1$, its relative velocity $\dot{s}_i(t)=v_{i-1}(t)-v_i(t)$, and its own velocity $v_i(t)$. A typical form is~\cite{orosz2010traffic}
\begin{equation}\label{Eq:HDVModel}
\dot{v}_i(t)=F\left(s_i(t),\dot{s}_i(t),v_i(t)\right), \quad i \in \Omega \backslash S,
\end{equation}
where $F(\cdot)$ is a nonlinear function.  
{ Both OVM and IDM can be written in this general form. Here, we use the OVM model to exemplify the HDVs' car-following behavior in~\eqref{Eq:HDVModel}, which has been widely considered in~\cite{jin2017optimal,huang2020learning,di2019cooperative,wang2020controllability}.} In OVM, the dynamics~\eqref{Eq:HDVModel} are
\begin{equation} \label{Eq:OVMmodel}
\dot{v}_i(t)=\alpha\left(v_{\mathrm{des}}\left(s_i(t)\right)-v_i(t)\right)+\beta\dot{s}_i(t),\quad i \in \Omega \backslash S,
\end{equation}
where $\alpha, \beta > 0$ denote the driver's sensitivity coefficients, and $v_{\mathrm{des}}(s)$ represents the spacing-dependent desired velocity of the human driver, given by a continuous piece-wise function
\begin{equation} \label{Eq:OVMDesiredVelocity}
v_{\mathrm{des}}(s)=\begin{cases}
0, &s\le s_{\mathrm{st}};\\
f_v(s), &s_{\mathrm{st}}<s<s_{\mathrm{go}};\\
v_{\max}, &s\ge s_{\mathrm{go}}.
\end{cases}
\end{equation}
In~\eqref{Eq:OVMDesiredVelocity}, the desired velocity $v_{\mathrm{des}}(s)$ becomes zero for a small spacing $s_{\mathrm{st}}$, and reaches a maximum value $v_{\max}$ for a large spacing $s_{\mathrm{go}}$. When $s_{\mathrm{st}}<s<s_{\mathrm{go}}$, the desired velocity is given by a monotonically increasing function $f_v (s)$, one typical choice of which is
\begin{equation}
\label{Eq:DesiredVelocityPolicy}
f_v(s) = \frac{v_{\max }}{2}\left(1-\cos (\pi\frac{s-s_{\mathrm{st}}}{s_{\mathrm{go}}-s_{\mathrm{st}}})\right).
\end{equation}

{ In the following, we proceed to use the general form~\eqref{Eq:HDVModel} of the car-following model to present the parametric system modeling and controllability/observability analysis.}

\subsection{Input/Output of Mixed Traffic System}

We now present the state, output and input vectors of the mixed traffic system shown in Fig.~\ref{Fig:SystemSchematic}.


\vspace{1mm}
\noindent\textbf{Equilibrium traffic state:} In an equilibrium traffic state, each vehicle moves with the same  velocity $v^{*}$ and the corresponding  spacing $s^{*}$. When each vehicle follows its predecessor, as shown in Fig.~\ref{Fig:SystemSchematic}, the equilibrium velocity of the traffic system is determined by the steady-state velocity of the head vehicle, indexed as $0$. If the head vehicle maintains a constant velocity $v_{0}$, we have $v^*=v_{0}$ for all other vehicles in Fig.~\ref{Fig:SystemSchematic}.

{ On the other hand, the equilibrium spacing for each vehicle might be heterogeneous\footnote{ We keep $s^*$ instead of a heterogeneous symbol $s_i^*,\,i\in\{1,2,\ldots,n\}$ for notational simplicity. Our methodology and results are directly applicable in the heterogeneous case. }} and can be non-trivial to obtain. If the HDVs' car-following dynamics~\eqref{Eq:HDVModel} are explicitly known, we can obtain the equilibrium spacing via solving
\begin{equation} \label{Eq:Equilibrium}
F\left(s^*,0,v^*\right)=0,
\end{equation}
which provides  equilibrium points $(s^*,v^*) $. However, $s^*$ becomes unknown if~\eqref{Eq:HDVModel} is not known accurately.  The equilibrium spacing for each CAV is a pre-designed variable~\cite{zheng2020smoothing}.

\vspace{1mm}
\noindent\textbf{System state:}
Assuming that the mixed traffic flow is moving around an equilibrium state $(s^*,v^*) $, we define the error state between actual and equilibrium point as ($i \in \Omega$)
\begin{equation} \label{Eq:StateDefinition}
\tilde{s}_i(t)=s_i(t)-s^*,\quad \tilde{v}_i(t)=v_i(t)-v^*,
\end{equation}
where $\tilde{s}_{i}(t)$, $\tilde{v}_{i}(t)$ represent the spacing error and velocity error of vehicle $i$ at time $t$, respectively. 
The error states of all the vehicles are then lumped as the mixed traffic system state $x(t)\in \mathbb{R}^{2n}$, given by
\begin{equation}\label{Eq:SystemState}
x(t)=\begin{bmatrix}\tilde{s}_{1}(t),\tilde{v}_{1}(t),\tilde{s}_{2}(t),\tilde{v}_{2}(t),\ldots,\tilde{s}_{n}(t),\tilde{v}_{n}(t)\end{bmatrix}^{\top}.
\end{equation}

\vspace{1mm}
\noindent\textbf{System output:}
Not all the variables in mixed traffic state $x(t)$ can be measured. As discussed above, the equilibrium spacing $s^*$ for the HDVs is non-trivial to get accurately due to unknown car-following dynamics~\eqref{Eq:HDVModel}. It is thus impractical to observe the spacing errors of the HDVs, \ie, $\tilde{s}_i(t)$ $(i \notin S)$. For the CAVs, their equilibrium spacing can be designed~\cite{zheng2020smoothing}, and thus their spacing error signal can be measured. 

We thus introduce the following output signal 
\begin{equation}\label{Eq:SystemOutput}
y(t)\!=\!\begin{bmatrix}\tilde{v}_{1}(t),\tilde{v}_{2}(t),\ldots,\tilde{v}_{n}(t),\tilde{s}_{i_1}(t),\tilde{s}_{i_2}(t),\ldots,\tilde{s}_{i_m}(t)\end{bmatrix}^{\top},
\end{equation}
where $y(t)\in \mathbb{R}^{n+m}$ consists of all measurable data, including the velocity errors of both the HDVs and the CAVs, \ie, $\tilde{v}_i(t)$ $(i \in \Omega)$, and the spacing errors of all the CAVs, \ie, $\tilde{s}_i(t)$ $(i \in S)$. 
The measurable output data are also marked in Fig.~\ref{Fig:SystemSchematic}, with velocity errors and spacing errors represented by black dashed arrows and blue dashed arrows, respectively.

\vspace{1mm}
\noindent\textbf{System input:} In mixed traffic flow, the HDVs are controlled by human drivers, while the CAVs' behavior can be designed. 
As used in~\cite{jin2017optimal,huang2020learning,di2019cooperative,wang2020controllability,zheng2020smoothing}, the acceleration of each CAV is assumed to be directly controlled 
\begin{equation} \label{Eq:CAVModel}
	\dot{v}_i(t)=u_i(t), \quad i \in S,
\end{equation}
where $u_i(t)$ is the control input of the CAV indexed as $i$. The acceleration signals of all the CAVs are lumped as the aggregate control input $u(t)\in\mathbb{R}^{m}$, given by
\begin{equation} \label{Eq:ControlInput}
	u(t) = \begin{bmatrix}u_{i_{1}}(t), u_{i_{2}}(t), \ldots, u_{i_{m}}(t)\end{bmatrix}^{\top}.
\end{equation}

In addition to the control input, we introduce an external input signal $\epsilon(t)\in\mathbb{R}$ of the mixed traffic system, which is defined as the velocity error of the head vehicle, given by
\begin{equation} \label{Eq:ExternalInput}
    \epsilon(t)=\tilde{v}_{0}(t)=v_0(t)-v^*.
\end{equation}
This external input signal plays a critical role in our subsequent system analysis and \method{DeeP-LCC} design. Since the head vehicle is also under human control, this input cannot be designed directly, but its past value can be measured and future value can be estimated.

\subsection{Linearized State-Space Model of Mixed Traffic System}
\label{Section:TrafficModel}

After specifying the system state, input and output, we now present a linearized mixed traffic model. Using~\eqref{Eq:Equilibrium} and applying the first-order Taylor expansion to~\eqref{Eq:HDVModel}, we obtain the following linearized model for each HDV 
\begin{equation}\label{Eq:LinearHDVModel}
\begin{cases}
\dot{\tilde{s}}_i(t)=\tilde{v}_{i-1}(t)-\tilde{v}_i(t),\\
\dot{\tilde{v}}_i(t)=\alpha_{1}\tilde{s}_i(t)-\alpha_{2}\tilde{v}_i(t)+\alpha_{3}\tilde{v}_{i-1}(t),\\
\end{cases} \, i \in \Omega \backslash S,
\end{equation}
where $\alpha_{1} = \frac{\partial F}{\partial s}, \alpha_{2} = \frac{\partial F}{\partial \dot{s}} - \frac{\partial F}{\partial v}, \alpha_{3} = \frac{\partial F}{\partial \dot{s}}$ with the partial derivatives evaluated at the equilibrium state ($s^*,v^*$). To reflect asymptotically stable driving behaviors of human drivers, we have $\alpha_{1}>0$, $\alpha_{2}>\alpha_{3}>0$ \cite{jin2017optimal}. Taking the OVM model~\eqref{Eq:OVMmodel} for example, the equilibrium equation~\eqref{Eq:Equilibrium} is given by
\begin{equation} \label{Eq:EquilibriumEquation_OVM}
	v_{\mathrm{des}}(s^*) = v^*,
\end{equation}
and the coefficients in the linearized dynamics~\eqref{Eq:LinearHDVModel} become
\begin{equation*}
\alpha_1 = \alpha \dot{v}_{\mathrm{des}}(s^*),\; \alpha_2=\alpha+\beta, \; \alpha_3=\beta,
\end{equation*}
where $\dot{v}_{\mathrm{des}}(s^*)$ denotes the derivative of $v_{\mathrm{des}}(s)$ at the equilibrium spacing $s^*$.

For the CAV,  we consider a second-order model 
\begin{equation} \label{Eq:LinearCAVModel}
	\begin{cases}
	\dot{\tilde{s}}_i(t)=\tilde{v}_{i-1}(t)-\tilde{v}_i(t),\\
	\dot{\tilde{v}}_i(t)=u_i(t),\\
	\end{cases} \quad i\in S.
\end{equation}
Based on the state, output and input vectors in~\eqref{Eq:StateDefinition}-\eqref{Eq:ExternalInput}, the linearized HDVs' car-following model~\eqref{Eq:LinearHDVModel} and the CAV's dynamics~\eqref{Eq:LinearCAVModel}, we derive a linearized state-space model of the mixed traffic in Fig.~\ref{Fig:SystemSchematic} as 
\begin{equation} \label{Eq:LinearSystemModel}
\begin{cases}
\dot{x}(t)=Ax(t)+Bu(t)+H\epsilon(t),\\
y(t)=Cx(t).
\end{cases}
\end{equation}
In~\eqref{Eq:LinearSystemModel}, the matrices $A \in \mathbb{R}^{2n\times 2n}, B \in \mathbb{R}^{2n \times m}, H\in \mathbb{R}^{2n \times 1},C\in \mathbb{R}^{(n+m) \times 2n}$  are given by
\begin{align*}
A&=\begin{bmatrix} A_{1,1} & & & &   \\
A_{2,2} & A_{2,1} & &  &   \\
& \ddots& \ddots&  &  \\
& & A_{n-1,2}& A_{n-1,1} &  \\
& & &  A_{n,2}&A_{n,1}\\
\end{bmatrix} ,\\
B &= \begin{bmatrix}
\mathbb{e}_{2n}^{2i_{1}}, \mathbb{e}_{2n}^{2i_{2}}, \ldots, \mathbb{e}_{2n}^{2i_{m}}
\end{bmatrix},\quad
H = \begin{bmatrix}
h_{1}^{\top},h_{2}^{\top},\ldots,h_n^{\top}
\end{bmatrix}^{\top},\\
C&=\begin{bmatrix}
\mathbb{e}_{2n}^{2}, \mathbb{e}_{2n}^{4}, \ldots, \mathbb{e}_{2n}^{2n},\mathbb{e}_{2n}^{2i_1-1},\mathbb{e}_{2n}^{2i_2-1},\ldots,\mathbb{e}_{2n}^{2i_m-1}
\end{bmatrix}^\top,
\end{align*}
where\footnote{The system matrices $A,B,C$ are indeed set functions with respect to the value of $S$~\cite{li2022cooperative}. For simplicity, the symbol $S$ is neglected.}
\begin{align*}
A_{i,1} &= \begin{cases}
P_1, \; i\notin S;\\
S_1, \; i\in S;
\end{cases} \;
A_{i,2} = \begin{cases}
P_2, \; i\notin S;\\
S_2, \; i\in S;
\end{cases}\\
h_{1} &= \begin{bmatrix} 1 \\ \alpha_{3}\end{bmatrix},\;
h_j= \begin{bmatrix} 0 \\ 0 \end{bmatrix},\; j \in \{2,3,\ldots,n\},
\end{align*}
with
\begin{align*}
P_{1} \!=\! \begin{bmatrix} 0 & -1 \\ \alpha_{1} & -\alpha_{2} \end{bmatrix}\!,
P_{2} \!=\! \begin{bmatrix} 0 & 1 \\ 0 & \alpha_{3} \end{bmatrix}\!,
	S_1\! =\! \begin{bmatrix} 0 & -1 \\ 0 & 0 \end{bmatrix}\!,
S_2\! =\! \begin{bmatrix} 0 & 1 \\ 0 & 0 \end{bmatrix}.
\end{align*}


\begin{remark}[State-feedback versus output-feedback]
    Most existing work on CAV control relies on state feedback which assumes a known equilibrium spacing $s^*$ and requires the system state $x(t)$ in~\eqref{Eq:SystemState} (see, \eg, the model-based strategies~\cite{jin2017optimal,zheng2020smoothing,di2019cooperative} and the data-driven strategies~\cite{gao2016data,huang2020learning,lan2021data}). The output-feedback case has been less investigated (two notable exceptions are~\cite{wang2020controllability,mousavi2021synthesis}). In practice, the equilibrium spacing $s^*$ is unknown and might be time-varying. Hence, we introduce a measurable output in~\eqref{Eq:SystemOutput} that does not use the HDVs' spacing errors. Also, the CAVs' spacing errors  play a critical role in car-following safety, and they should be constrained for collision-free guarantees. The output-feedback and constraint requirements motivate us to use an MPC framework later. 
    \markend
\end{remark}

\begin{remark}[Unknown car-following behavior]
 One challenge for mixed traffic control lies in the unknown car-following behavior~\eqref{Eq:HDVModel}.  After linearization, the state-space model~\eqref{Eq:LinearSystemModel} of the mixed traffic system remains unknown. We focus on a data-driven predictive control method that directly relies on the driving data of HDVs. Before presenting the methodology, we need to investigate two fundamental control-theoretic properties of the mixed traffic system, controllability and observability, which are essential to establish  data-driven predictive control~\cite{coulson2019data}. Our previous work on LCC has investigated the special case with only one CAV~\cite{wang2021leading}. In the next section, we generalize these results to the case with possibly multiple CAVs and HDVs coexisting (see Fig.~\ref{Fig:SystemSchematic}).
 \markend
\end{remark}

\section{Controllability and Observability \\ of Mixed Traffic Systems}
\label{Sec:3}

Controllability and observability are two fundamental properties in dynamical systems~\cite{skogestad2007multivariable}. 
For mixed traffic systems, existing research~\cite{wang2021leading,jin2017optimal} has revealed the controllability for the scenario of one single CAV and multiple HDVs, \ie, $|S|=1$. These results have been unified in the recent LCC framework with one single CAV~\cite{wang2021leading}.

\begin{lemma}[\!\rm{\cite[Corollary 1]{wang2021leading}}] \label{Lemma:CFLCC_Controllability}
	When $S = \{1\}$, the linearized mixed traffic system~\eqref{Eq:LinearSystemModel} is controllable if we have
	\begin{equation}
	\vspace{2mm}
	\label{Eq:ControllabilityCondition}
	\alpha _{1}- \alpha _{2} \alpha _{3}+ \alpha _{3}^{2} \neq 0.
	\end{equation}
\end{lemma}

\begin{lemma}[\!\rm{\cite[Theorem 2]{wang2021leading}}] \label{Lemma:LCC_Controllability}
	When $S = \{i_1\}$ with $ 1 < i_1 \leq n$, the linearized mixed traffic system~\eqref{Eq:LinearSystemModel} is not controllable but is stabilizable, if~\eqref{Eq:ControllabilityCondition} holds. Moreover, if~\eqref{Eq:ControllabilityCondition} holds, the subsystem consisting of the states $\tilde{s}_{1},\tilde{v}_{1},\ldots,\tilde{s}_{i_1-1},\tilde{v}_{i_1-1}$ is not controllable but is stable, while the subsystem consisting of the states $\tilde{s}_{i_1},\tilde{v}_{i_1},\ldots,\tilde{s}_{n},\tilde{v}_{n}$ is controllable.
\end{lemma}

One physical interpretation of Lemmas~\ref{Lemma:CFLCC_Controllability} and~\ref{Lemma:LCC_Controllability} is that the control input of the single CAV has no influence on the state of its preceding HDVs, but has full control of the motion of its following HDVs, when~\eqref{Eq:ControllabilityCondition} holds. 

We now present the controllability properties of the general mixed traffic system with multiple CAVs and HDVs in Fig.~\ref{Fig:SystemSchematic}.

\begin{theorem}[Controllability]
	\label{Theorem:Controllability}
	Consider the mixed traffic system~\eqref{Eq:LinearSystemModel}, where there exist $m$ $(m \geq 1)$ CAVs with indices $S=\{i_1,i_2,\ldots,i_m\}$, $i_1 < i_2 < \ldots < i_m$. We have:
	\begin{enumerate}
		\item When $1 \in S$, \ie, $i_1 = 1$, the mixed traffic system is controllable if~\eqref{Eq:ControllabilityCondition} holds.
		\item When $1 \notin S$, \ie, $i_1 > 1$, the mixed traffic system is not controllable but is stabilizable, if~\eqref{Eq:ControllabilityCondition} holds. Particularly, when~\eqref{Eq:ControllabilityCondition} holds, the subsystem consisting of the states $\tilde{s}_{1},\tilde{v}_{1},\ldots,\tilde{s}_{i_1-1},\tilde{v}_{i_1-1}$ is not controllable but is stable, while the subsystem consisting of the states $\tilde{s}_{i_1},\tilde{v}_{i_1},\ldots,\tilde{s}_{n},\tilde{v}_{n}$ is controllable.
	\end{enumerate}
\end{theorem}

\begin{IEEEproof}
 The proof combines the controllability invariance after state feedback with Lemmas~\ref{Lemma:CFLCC_Controllability} and~\ref{Lemma:LCC_Controllability}. The details are not mathematically involved, and we provide them in Appendix~\ref{Appendix:Controllability} for completeness.
\end{IEEEproof}

This result indicates that the general mixed traffic system consisting of multiple CAVs and HDVs is not controllable (but stabilizable) unless the vehicle immediately behind the head vehicle is a CAV. This is expected, since the motion of the HDVs between the head vehicle and the first CAV (\ie, vehicles indexed from $1$ to $i_1-1$) can not be influenced by the CAVs' control inputs. 

We consider an output-feedback controller design. It is essential to evaluate the observability of the mixed traffic system~\eqref{Eq:LinearSystemModel}. The notion of observability quantifies the ability of reconstructing the system state from its output measurements. 
By adapting~\cite[Theorem 4]{wang2021leading}, we have the following result. 

\begin{theorem}[Observability]
	\label{Theorem:Observability}
	The general mixed traffic system given by~\eqref{Eq:LinearSystemModel}, where there exist $m$ $(m \geq 1)$ CAVs, is observable when~\eqref{Eq:ControllabilityCondition} holds.
\end{theorem}


The slight asymmetry between Theorems~\ref{Theorem:Controllability} and~\ref{Theorem:Observability} is due to the fact that the control input~\eqref{Eq:ControlInput} only includes the CAVs' acceleration, while the system output~\eqref{Eq:SystemOutput} consists of the velocity error of all the vehicles and the spacing error of the CAVs. Theorem~\ref{Theorem:Observability} reveals the observability of the full state $x(t)$ in mixed traffic under a mild condition. This observability result facilitates the design of our \method{DeeP-LCC} strategy, which will be detailed in the next two sections. 

\section{Data-EnablEd Predictive Control}

\label{Sec:4}

In this section, we give an overview of the data-driven methodology on non-parametric representation of system behavior and Data-Enabled Predicted Control (DeePC); more details can be referred to~\cite{coulson2019data,dorfler2022bridging}. 

\subsection{Non-Parametric Representation of System Behavior}

DeePC works on discrete-time systems \cite{coulson2019data}. Let us consider a discrete-time LTI system
\begin{equation} \label{Eq:LTI}
	\begin{cases}
	x(k+1) = A_{\mathrm{d}}x(k) + B_{\mathrm{d}}u(k),\\
	y(k) = C_{\mathrm{d}}x(k) + D_{\mathrm{d}}u(k),
	\end{cases}
\end{equation}
where $A_{\mathrm{d}} \in \mathbb{R}^{n \times n}$, $B_{\mathrm{d}} \in \mathbb{R}^{n \times m}$, $C_{\mathrm{d}} \in \mathbb{R}^{p \times n}$, $D_{\mathrm{d}} \in \mathbb{R}^{p \times m}$, and $x(k)\in \mathbb{R}^n,\,u(k) \in \mathbb{R}^m,\,y(k)\in \mathbb{R}^p$ denotes the internal state, control input, and output at time $k$ ($k \in \mathbb{N}$), 
respectively. By slight abuse of notation, we use the symbols $n,m,p$ to denote system dimensions only in this section.

Classical control strategies typically follow sequential system identification and model-based controller design. They rely on the explicit system model  $A_\mathrm{d},B_\mathrm{d},C_\mathrm{d},D_\mathrm{d}$ in~\eqref{Eq:LTI}. 
One typical strategy is the celebrated MPC framework~\cite{camacho2013model}. The performance of MPC is closely related to the accuracy of the system model. Although many system identification methods are available~\cite{ljung2017system}, it is still non-trivial to obtain an accurate model for complex systems, \eg, the mixed traffic system with complex nonlinear human driving behavior. 
The  recent DeePC~\cite{coulson2019data} is a \textit{non-parametric method} that bypasses system identification and directly designs the control input compatible with historical data. In particular,  DeePC directly uses historical data to predict the system behavior based on Willems' \emph{fundamental lemma}~\cite{willems2005note}. 

\begin{definition} \label{Def:PersistentExcitation}
	The signal  $\omega = \col \left(\omega(1),\omega(2),\ldots,\omega(T)\right)$ of length $T$ $(T\in \mathbb{N})$ is persistently exciting of order $l$ $(l \leq T,\,l\in \mathbb{N})$ if the following Hankel matrix
	\begin{equation}
		\mathcal{H}_{l}(\omega):=\begin{bmatrix}
			\omega(1) &\omega(2) & \cdots & \omega(T-l+1) \\
			\omega(2) &\omega(3) & \cdots & \omega(T-l+2) \\
			\vdots & \vdots & \ddots & \vdots \\
			\omega(l) &\omega(l+1) & \cdots & \omega(T)
		\end{bmatrix},
	\end{equation}
	is of full row rank.
\end{definition}

The Williem's fundamental lemma begins by collecting a length-$T$ ($T\in \mathbb{N}$) sequence of trajectory data from system~\eqref{Eq:LTI}, consisting of the input sequence ${u}^\mathrm{d}=\col ({u}^\mathrm{d}(1),\ldots,$ ${u}^\mathrm{d}(T))\in \mathbb{R}^{mT}$ and the corresponding output sequence $y ^\mathrm{d}=\col (y^\mathrm{d}(1),\ldots,y^\mathrm{d}(T)) \in \mathbb{R}^{pT}$. Then, it aims to utilize this pre-collected length-$T$ trajectory to directly construct valid length-$L$ ($L\in \mathbb{N}$) trajectories of the system, consisting of input sequence $u^\mathrm{s} \in \mathbb{R}^{mL}$ and output sequence $y^\mathrm{s}\in \mathbb{R}^{pL}$. 
\begin{lemma}[Fundamental Lemma~\cite{willems2005note}]
	\label{Lemma:FundamentalLemma}
	Consider a controllable LTI system~\eqref{Eq:LTI} and assume the input sequence $u^\mathrm{d}$ to be persistently exciting of order $L+n$. Then, $(u^\mathrm{s},y^\mathrm{s})$ is a length-$L$ input/output trajectory of system~\eqref{Eq:LTI} if and only if
	there exists $g \in \mathbb{R}^{T-L+1}$ such that
	\begin{equation}
	\label{Eq:FundamentalLemma}
		\begin{bmatrix}
		\mathcal{H}_L(u^\mathrm{d}) \\ \mathcal{H}_L(y^\mathrm{d})
		\end{bmatrix}g=
		\begin{bmatrix}
		u^\mathrm{s} \\ y^\mathrm{s}
		\end{bmatrix}.
	\end{equation}
\end{lemma}

This fundamental lemma reveals that given a controllable LTI system, 
the subspace consisting of all valid length-$L$ trajectories is identical to the range space of the Hankel matrix of depth $L$ generated by a sufficiently rich input signal. Rather than identifying a parametric model, this lemma allows for non-parametric representation of system behaviors. 

\subsection{Data-EnablEd Predictive Control}

Define $T_\mathrm{ini} \in \mathbb{N}$, $N \in \mathbb{N}$ as the time length of ``past data" and ``future data", respectively. The data Hankel matrices constructed from the pre-collected data $(u^\mathrm{d},y^\mathrm{d})$ are partitioned into the two parts (corresponding to past data and future data):  
\begin{equation}
\label{Eq:DataHankel}
\begin{bmatrix}
U_{\mathrm{p}} \\
U_{\mathrm{f}}
\end{bmatrix}:=\mathcal{H}_{T_{\mathrm{ini}}+N}(u^{\mathrm{d}}),\;
\begin{bmatrix}
Y_{\mathrm{p}} \\
Y_{\mathrm{f}}
\end{bmatrix}:=\mathcal{H}_{T_{\mathrm{ini}}+N}(y^{\mathrm{d}}),
\end{equation}
where $U_{\mathrm{p}}$ and $U_{\mathrm{f}}$ consist of the first $T_{\mathrm{ini}}$ block rows and the last $N$ block rows of $\mathcal{H}_{T_{\mathrm{ini}}+N}(u^{\mathrm{d}})$, respectively (similarly for $Y_{\mathrm{p}}$ and $Y_{\mathrm{f}}$). The same column in $\col(U_{\mathrm{p}},U_{\mathrm{f}})$ and $\col(Y_{\mathrm{p}},Y_{\mathrm{f}})$ represents the ``past'' input/output signal of length $T_{\mathrm{ini}}$  and the ``future'' input/output signal of length $N$ within a length-$(T_{\mathrm{ini}}+N)$ trajectory of~\eqref{Eq:LTI}.

\begin{figure*}[t]
\setlength{\abovecaptionskip}{0pt}
	\centering
	\includegraphics[scale=0.43]{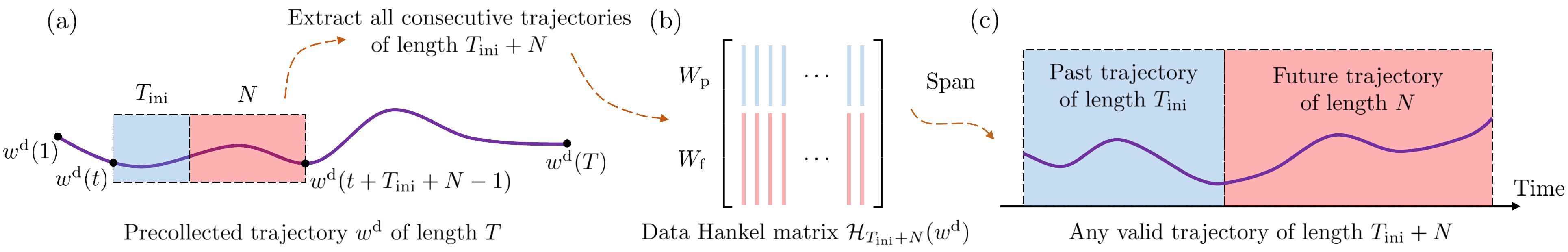}
	\caption{ Interpretation of the fundamental lemma in~\eqref{Eq:DeePCAchievability}. Here, we use $w$ to denote input/output trajectory pair $(u,y)$. (a) A consecutive length-$T$ trajectory is collected ${w}^\mathrm{d}=\col ({w}^\mathrm{d}(1),\ldots,$ ${w}^\mathrm{d}(T))$. (b) All consecutive length-$(T_\mathrm{ini}+N)$ trajectories are extracted to construct the data Hankel matrix $\mathcal{H}_{T_\mathrm{ini}+N}(w^\mathrm{d})$. Particularly, each trajectory is partitioned into two parts, past data of length $T_\mathrm{ini}$ colored in blue and future data of length $T$ colored in red. (c) The range space of this data Hankel matrix contains all valid length-$(T_\mathrm{ini}+N)$ trajectories of the underlying system.}
	\label{Fig:FundamentalLemma}
\end{figure*}

At time step $t$, we define 
$
    u_{\mathrm{ini}}=\col(u(t-T_{\mathrm{ini}}),u(t-T_{\mathrm{ini}}+1),\ldots,u(t-1)),
u= \col(u(t),u(t+1),\ldots,u(t+N-1))
$
as the past control input sequence with time length $T_{\mathrm{ini}}$, and the future control input sequence with time horizon $N$, respectively (similarly for $y_\mathrm{ini},y$).
Then, we have the following proposition, which is a reformulation of Lemma~\ref{Lemma:FundamentalLemma}.

\begin{proposition}[\!\!\!\cite{markovsky2008data}] \label{Proposition:ReformulatedFundamentalLemma}
    Consider a controllable LTI system~\eqref{Eq:LTI} and assume the input sequence $u^\mathrm{d}$ to be persistently exciting of order $T_{\mathrm{ini}}+N+n$. Then, $\col (u_\mathrm{ini},u,y_\mathrm{ini},y)$ is a length-$(T_{\mathrm{ini}}+N)$ input/output trajectory of system~\eqref{Eq:LTI} if and only if
	there exists $g \in \mathbb{R}^{T-T_\mathrm{ini}-N+1}$ such that
\begin{equation}
\label{Eq:DeePCAchievability}
\begin{bmatrix}
U_\mathrm{p} \\ Y_\mathrm{p} \\ U_\mathrm{f} \\ Y_\mathrm{f}
\end{bmatrix}g=
\begin{bmatrix}
u_\mathrm{ini} \\ y_\mathrm{ini} \\ u \\ y
\end{bmatrix}.
\end{equation}
In particular, if $T_{\mathrm{ini}} \geq \nu $, where $\nu$ denotes the lag\footnote{The lag $ \nu $ of a system $(A,B,C,D)$ is the smallest integer such that the observability matrix $\col \left(C, C A, \ldots, C A^{\nu-1}\right)$ has full column rank.} of system~\eqref{Eq:LTI}, $y$ is unique from~\eqref{Eq:DeePCAchievability}, $\forall u_\mathrm{ini}, y_\mathrm{ini}, u$.
\end{proposition}

A schematic of Proposition \ref{Proposition:ReformulatedFundamentalLemma} is shown in Fig.~\ref{Fig:FundamentalLemma}. The formulation~\eqref{Eq:DeePCAchievability} indicates that given a past input/output trajectory $(u_\mathrm{ini},y_\mathrm{ini})$, one can predict the future output sequence $y$ under a future input sequence $u$ directly from pre-collected data $(u^\mathrm{d},y^\mathrm{d})$.   
It is known that when $T_{\mathrm{ini}} \geq \nu $, one can estimate the initial state based on model~\eqref{Eq:LTI} and the past input/output trajectory $(u_\mathrm{ini},y_\mathrm{ini})$. Thus,~\eqref{Eq:DeePCAchievability} implicitly estimates the initial state to predict the future trajectory $\col(u,y)$ without an explicit parametric model~\cite{coulson2019data}.

\begin{algorithm}[t]
	\caption{DeePC \cite{coulson2019data}}
	\label{Alg:DeePC}
	\begin{algorithmic}[1]
		\Require
		Offline data $(u^{\mathrm{d}},y^{\mathrm{d}})$, initial time $t_0$, final time $t_f$;
		\State Construct data Hankel matrices $U_\mathrm{p} , U_\mathrm{f},  Y_\mathrm{p} , Y_\mathrm{f}$;
		\State Initialize past data $(u_\mathrm{ini},y_\mathrm{ini})$ before time $t_0$;
		\While{$t_0 \leq t \leq t_f$}
		\State Solve~\eqref{Eq:DeePC} to get an optimal input sequence $u^*=\col(u^*(t),u^*(t+1),\ldots,u^*(t+N-1)$;
		\State Apply the input $u(t) \leftarrow u^*(t)$;
		\State $t \leftarrow t+1$ and update past input/output data $(u_{\mathrm{ini}},y_{\mathrm{ini}})$;
		\EndWhile
	\end{algorithmic}
\end{algorithm}

At each time step $t$, DeePC relies on the data-centric representation~\eqref{Eq:DeePCAchievability} to predict future system behavior and solves the following optimization problem \cite{coulson2019data}
\begin{equation} \label{Eq:DeePC} 
\begin{aligned}
\min_{g,u,y} \quad &J(y,u)\\
\st \quad &\eqref{Eq:DeePCAchievability},\,u \in \mathcal{U}, \, y \in \mathcal{Y},
\end{aligned}
\end{equation}
where $J(y,u)$ denotes the control objective function, and $u\in\mathcal{U},y\in\mathcal{Y}$ represents the input/output constraints, \eg, safety guarantees and control saturation. Problem~\eqref{Eq:DeePC} is solved in a receding horizon manner (see Algorithm~\ref{Alg:DeePC}). %
For comparison, we also present a standard output-feedback MPC
\begin{equation} \label{Eq:MPC} 
\begin{aligned}
\min_{u} \quad &J(y,u)\\
\st \quad & x(t) = \hat{x}(t),\\
&\eqref{Eq:LTI},\,\forall k\in \{t,t+1,\ldots,t+N-1\},\\
&u \in \mathcal{U}, \, y \in \mathcal{Y},
\end{aligned}
\end{equation}
where $\hat{x}(t)$ denotes the estimated initial state at time $t$. 
%

Despite its well-recognized effectiveness, one crucial challenge for the standard MPC \eqref{Eq:MPC} is the requirement of an explicit parametric model~\eqref{Eq:LTI}, which is necessary in estimating the initial state $\hat{x}(t)$ and predicting future system behaviors. By contrast, DeePC~\eqref{Eq:DeePC} focuses on the data-centric non-parametric representation and bypasses the state estimation procedure~\cite{coulson2019data}. 
Recent work has revealed the equivalence between DeePC and sequential system identification and MPC for discrete-time LTI systems under mild conditions, and comparable performance of DeePC with respect to MPC based on accurate model knowledge in applications to nonlinear and non-deterministic systems~\cite{dorfler2022bridging}.

\section{\method{DeeP-LCC} for Mixed Traffic Flow}

\label{Sec:5}

	The Willems' fundamental lemma requires the controllability of the discrete-time LTI system~\eqref{Eq:LTI} and the persistent excitation of pre-collected input data $u^\mathrm{d}$~\cite{willems2005note}. 
	As shown in Theorem~\ref{Theorem:Controllability}, the mixed traffic system is not always controllable, and thus the original DeePC cannot be directly applied for mixed traffic control. 
	
	In this section, we introduce an external input signal for mixed traffic. Together with original control input, this leads to controllability. 
%
We first reformulate mixed traffic model~\eqref{Eq:LinearSystemModel}, and then present \method{DeeP-LCC} for mixed traffic control.

\subsection{Model Reformulation with External Input}

\label{Sec:SystemReformulation}

Theorem~\ref{Theorem:Controllability} has revealed that the mixed traffic system~\eqref{Eq:LinearSystemModel} is not controllable when $1 \notin S$, \ie, the first vehicle behind the head vehicle is not a CAV. Still, controllability is a desired property, which is required in Willems' fundamental lemma (Lemma~\ref{Lemma:FundamentalLemma}) to guarantee the data-centric behavior representation. To resolve this, we introduce a variant of the original system~\eqref{Eq:LinearSystemModel} that is fully controllable. 

The velocity error of the head vehicle $ \epsilon(t)=v_0(t)-v^*$ is an external input in~\eqref{Eq:LinearSystemModel}. This signal is not directly controlled, but can be measured in practice. Define $
	\hat{u}(t) = \col \left(\epsilon (t),u (t) \right)
$
as a combined input signal and $\widehat{B} = \begin{bmatrix} H,B
\end{bmatrix}$ as the corresponding input matrix. The model for the mixed traffic system becomes
\begin{equation} \label{Eq:TransformedMixedTrafficSystem}
	\begin{cases} 
\dot{x}(t)=Ax(t)+\widehat{B} \hat{u}(t),\\
y(t)=Cx(t),
\end{cases}
\end{equation}
for which we have the following result.
\begin{corollary}[Controllability and Observability of the Reformulated Traffic Model]
	\label{Corollary:TransformedSystemControllability}
 Suppose there exist $m$ $(m \geq 1)$ CAVs. Then, system~\eqref{Eq:TransformedMixedTrafficSystem} is controllable and observable if~\eqref{Eq:ControllabilityCondition} holds.
\end{corollary}

The proof is similar to that of the system when the first vehicle behind the head vehicle is a CAV~\cite{wang2021leading}, \ie, $1 \in S$; we refer the interested reader to~\cite[Corollary 1]{wang2021leading}. For observability, it is immediate to see that system~\eqref{Eq:TransformedMixedTrafficSystem} shares the same output dynamics as system~\eqref{Eq:LinearSystemModel}, whose observability result has been proved in Theorem~\ref{Theorem:Observability}. 

By Corollary~\ref{Corollary:TransformedSystemControllability},  we can apply the fundamental lemma using a combined input $\hat{u}(t)$ consisting of the internal control input (\ie, the acceleration signals $u(t)$ of the CAVs) and the external input (\ie, the velocity error $\epsilon(t)$ of the head vehicle). For simplicity, we use the original system model~\eqref{Eq:LinearSystemModel} where the two input signals $u(t),\epsilon(t)$ are still separated. 
Finally, the system model~\eqref{Eq:LinearSystemModel} is in continuous-time domain. We transform it to the discrete-time domain 
\begin{equation} \label{Eq:DT_TrafficModel}
\begin{cases}
x(k+1) = A_\mathrm{d}x(k) + B_\mathrm{d}u(k) + H_\mathrm{d} \epsilon(k),\\
y(k) = C_\mathrm{d}x(k),
\end{cases}
\end{equation}
where $
A_\mathrm{d} = e^{A\Delta t}\in \mathbb{R}^{2n\times 2n},B_\mathrm{d}=\int_{0}^{\Delta t} e^{A t} Bd t\in \mathbb{R}^{2n\times m},H_\mathrm{d}=\int_{0}^{\Delta t} e^{A t} Hd t \in \mathbb{R}^{2n\times 1},C_\mathrm{d}=C \in \mathbb{R}^{(n+m)\times 2n} 
$, and $\Delta t>0$ is the sampling time interval. 

{ 
\begin{assumption}
\label{Assumption:PreserveControllability}
Denote $\lambda_i,\,i=1,\ldots,2n$ as the eigenvalues of $A$ in the continuous-time mixed traffic system model~\eqref{Eq:LinearSystemModel}. We have $\left|\operatorname{Im}\left[\lambda_i-\lambda_j\right]\right| \neq 2 \pi k/\Delta t, \; k=1,2,\ldots $, whenever  $\operatorname{Re}\left[\lambda_i-\lambda_j\right]=0,\;i,j=1,\ldots,2n$.
\end{assumption}

As revealed in~\cite[Theorem 6.9]{chen1984linear}, Assumption~\ref{Assumption:PreserveControllability} is a sufficient condition to preserve controllability and observability after discretization from~\eqref{Eq:LinearSystemModel} to the discrete-time system model~\eqref{Eq:DT_TrafficModel}.}

\subsection{Non-Parametric Representation of Mixed Traffic Behavior}

\label{Sec:DataCollection}


\noindent\textbf{Data collection:} 
 We begin by collecting a length-$T$ trajectory data from the mixed traffic system shown in Fig.~\ref{Fig:SystemSchematic}. Precisely, the collected data includes: 
 \begin{enumerate}
     \item the combined input sequence $\hat{u}^\mathrm{d}=\col (\hat{u}^\mathrm{d}(1),\ldots,$ $\hat{u}^\mathrm{d}(T))\in \mathbb{R}^{(m+1)T}$, consisting of CAVs' acceleration sequence $u^\mathrm{d}=\col (u^\mathrm{d}(1),\ldots,$ $u^\mathrm{d}(T))\in \mathbb{R}^{mT}$ and the velocity error sequence of the head vehicle $\epsilon ^\mathrm{d}=\col (\epsilon^\mathrm{d}(1),\ldots,\epsilon^\mathrm{d}(T)) \in \mathbb{R}^{T}$;
     \item the corresponding output sequence of the mixed traffic system $y ^\mathrm{d}=\col (y^\mathrm{d}(1),\ldots,y^\mathrm{d}(T)) \in \mathbb{R}^{(n+m)T}$. 
 \end{enumerate}

The pre-collected data are then partitioned into two parts, corresponding to ``past data'' of length $T_{\mathrm{ini}}$ and ``future data'' of length $N$. Precisely, define 
\begin{equation}
\begin{gathered}
\label{Eq:DataHankel}
\begin{bmatrix}
U_{\mathrm{p}} \\
U_{\mathrm{f}}
\end{bmatrix}:=\mathcal{H}_{T_{\mathrm{ini}}+N}(u^{\mathrm{d}}), \quad \begin{bmatrix}
E_{\mathrm{p}} \\
E_{\mathrm{f}}
\end{bmatrix}:=\mathcal{H}_{T_{\mathrm{ini}}+N}(\epsilon^{\mathrm{d}}),\\
\begin{bmatrix}
Y_{\mathrm{p}} \\
Y_{\mathrm{f}}
\end{bmatrix}:=\mathcal{H}_{T_{\mathrm{ini}}+N}(y^{\mathrm{d}}),
\end{gathered}    
\end{equation}
where $U_{\mathrm{p}}$ and $U_{\mathrm{f}}$ consist of the first $T_{\mathrm{ini}}$ block rows and the last $N$ block rows of $\mathcal{H}_{T_{\mathrm{ini}}+N}(u^{\mathrm{d}})$, respectively (similarly for $E_{\mathrm{p}}, E_{\mathrm{f}}$ and $Y_{\mathrm{p}}, Y_{\mathrm{f}}$).
 
These pre-collected data samples could be generated offline, or collected from the historical trajectories of those involved vehicles. According to Lemma~\ref{Lemma:FundamentalLemma}, the following assumption is needed for the pre-collected data (recall that the order of the mixed traffic system is $2n$).

\begin{assumption} \label{Assumption:PersistentExcitation}
 The combined input sequence $\hat{u}^\mathrm{d}$ is persistently exciting of order $T_{\mathrm{ini}}+N+2n$.\markend
\end{assumption}
 
{ Note that the external input, \ie, the velocity error of the head vehicle $\epsilon (t)$, is controlled by a human driver. Although it cannot be arbitrarily designed, it is always oscillating around zero since the driver always attempts to maintain the equilibrium velocity while suffering from small perturbations. Thus, given a  trajectory with length
 \begin{equation} \label{Eq:DataLength}
     T \geq (m+1)(T_{\mathrm{ini}}+N+2n)-1,
 \end{equation}
which allows for a $\hat{u}^\mathrm{d}$ Hankel matrix of order $T_{\mathrm{ini}}+N+2n$ to have a larger column number than the row number, and persistently exciting acceleration input $u(t)$ of the CAVs (\eg, i.i.d. noise with zero mean), the persistent excitation in Assumption~\ref{Assumption:PersistentExcitation} is naturally satisfied. }

 \vspace{1mm}
\noindent\textbf{Behavior Representation:} 
Similar to Proposition~\ref{Proposition:ReformulatedFundamentalLemma}, we have the following result: at time step $t$, define
\begin{equation}
\begin{aligned}
u_{\mathrm{ini}}&=\col\left(u(t-T_{\mathrm{ini}}),u(t-T_{\mathrm{ini}}+1),\ldots,u(t-1)\right),\\
u&= \col\left(u(t),u(t+1),\ldots,u(t+N-1)\right),
\end{aligned}    
\end{equation}
as the control sequence within a past time length $T_{\mathrm{ini}}$, and the control sequence within a predictive time length $N$, respectively (similarly for $\epsilon_\mathrm{ini},\epsilon$ and $y_\mathrm{ini},y$).


\begin{proposition} \label{Proposition:DeePLCC_FundamentalLemma}
Suppose~\eqref{Eq:ControllabilityCondition} and Assumptions~\ref{Assumption:PreserveControllability} and~\ref{Assumption:PersistentExcitation} hold. Any length-$(T_{\mathrm{ini}}+N)$ trajectory of the mixed traffic system~\eqref{Eq:DT_TrafficModel}, denoted as $\col (u_\mathrm{ini},\epsilon_\mathrm{ini},y_\mathrm{ini},u,\epsilon,y),$ 
can be constructed via
\begin{equation}
\label{Eq:AdaptedDeePCAchievability}
\begin{bmatrix}
U_\mathrm{p} \\ E_\mathrm{p}\\Y_\mathrm{p} \\ U_\mathrm{f} \\ E_\mathrm{f}\\ Y_\mathrm{f}
\end{bmatrix}g=
\begin{bmatrix}
u_\mathrm{ini} \\ \epsilon_\mathrm{ini}\\ y_\mathrm{ini} \\ u \\\epsilon \\ y
\end{bmatrix},
\end{equation}
where $g\in \mathbb{R}^{T-T_\mathrm{ini}-N+1}$. If $T_{\mathrm{ini}} \geq 2n $, $y$ is unique from~\eqref{Eq:AdaptedDeePCAchievability}, $\forall u_\mathrm{ini} ,\epsilon_\mathrm{ini}, y_\mathrm{ini},u,\epsilon$. 
\end{proposition}

\begin{IEEEproof}
Condition~\eqref{Eq:ControllabilityCondition} and Assumption~\ref{Assumption:PreserveControllability} guarantee the controllability and observability of the mixed traffic system~\eqref{Eq:DT_TrafficModel}, and Assumption~\ref{Assumption:PersistentExcitation} offers the persistent excitation property of pre-collected data. Then, this result can be  derived from Proposition~\ref{Proposition:ReformulatedFundamentalLemma}. Since the mixed traffic system is observable under condition~\eqref{Eq:ControllabilityCondition}, its lag is not larger than its state dimension $2n$, and thus we have the uniqueness of $y$ by Proposition~\ref{Proposition:ReformulatedFundamentalLemma}.
\end{IEEEproof}

Proposition~\ref{Proposition:DeePLCC_FundamentalLemma} reveals that by collecting traffic data, one can directly predict the future trajectory of the mixed traffic system. 
We thus require no explicit model of HDVs' car-following behavior. Note that HDVs are controlled by human drivers and have complex and uncertain dynamics. 
This result allows us to bypass a parametric system model and directly use non-parametric data-centric representation for the behavior of the mixed traffic system.

\subsection{Design of Cost Function and Constraints in \method{DeeP-LCC}}

Motivated by DeePC \eqref{Eq:DeePC}, 
we show how to utilize the non-parametric behavior representation~\eqref{Eq:AdaptedDeePCAchievability} to design the control input of the CAVs. We design the future behavior $(u,\epsilon,y)$ for the mixed traffic system in a receding horizon manner. This is based on pre-collected data $(u^\mathrm{d},\epsilon^\mathrm{d},y^\mathrm{d})$ and the most recent past data $(u_\mathrm{ini},\epsilon_\mathrm{ini},y_\mathrm{ini})$ that are updated online.

Compared to the standard DeePC~\eqref{Eq:DeePCAchievability}, one unique feature of~\eqref{Eq:AdaptedDeePCAchievability}   is the introduction of the external input sequence, \ie, the velocity error $\epsilon$ of the head vehicle. The past external input sequence $\epsilon_\mathrm{ini}$ can be collected in the control process, but the future external input sequence $\epsilon$ cannot be designed and is also unknown in practice. Although its future behavior might be predicted based on traffic conditions ahead, it is non-trivial to achieve an accurate prediction. Since the driver always attempts to maintain the equilibrium velocity, one~natural~approach is to assume that the future velocity error of the head vehicle is zero, \ie,
\begin{equation} \label{Eq:FutureExternalInput}
\epsilon = \mathbb{0}_N.
\end{equation}

Similar to LCC~\cite{wang2021leading}, we consider the performance of the entire mixed traffic system in Fig.~\ref{Fig:SystemSchematic} for controller design. Precisely, we use a quadratic cost function $J(y,u)$ to quantify the mixed traffic performance by penalizing the output deviation (recall that $y$ in~\eqref{Eq:SystemOutput} represents the measurable deviation from equilibrium) and the energy of control input $u$, defined as
\begin{equation} \label{Eq:CostDefinition}
J(y,u) = \sum\limits_{k=t}^{t+N-1}\left( \left\|y(k)\right\|_{Q}^{2}+\left\|u(k)\right\|_{R}^{2}\right),     
\end{equation}
where the weight matrices $Q$ and $R$ are set as 
$
Q=\diag(Q_v,Q_s)
$
with $Q_v = \diag(w_v,\ldots,w_v) \in \mathbb{R}^{n\times n}$, $Q_s = \diag(w_s,\ldots,w_s) \in \mathbb{R}^{m\times m}$ and $R = \diag(w_u,\ldots,w_u) \in \mathbb{R}^{m\times m} $ with $w_v,w_s,w_u$ representing the penalty for the velocity errors of all the vehicles, spacing errors of all the CAVs, and control inputs of the CAVs, respectively.

Now, we introduce several constraints for CAV control in mixed traffic. First, the safety constraint for collision-free guarantees need to be considered. To address this, we impose a lower bound on the spacing error of each CAV, given by
\begin{equation} \label{Eq:Constraint_MinimumSpacing}
	\tilde{s}_{i} \geq \tilde{s}_\mathrm{min}, \, i\in S,
\end{equation}
with $\tilde{s}_\mathrm{min}$ denoting the minimum spacing error for each CAV. With appropriate choice of $\tilde{s}_\mathrm{min}$, the rear-end collision of the CAVs is avoided whenever feasible.

Second, to attenuate traffic perturbations, existing CAVs controllers tend to leave an extremely large spacing from the preceding vehicle (see, \eg,~\cite{stern2018dissipation} and the discussions in~\cite[Section V-D]{wang2020controllability}), which in practice might cause vehicles from adjacent lanes to cut in. To tackle this problem, we introduce a maximum spacing constraint for each CAV, shown as
\begin{equation} \label{Eq:Constraint_MaximumSpacing}
	\tilde{s}_{i} \leq \tilde{s}_\mathrm{max}, \, i\in S,
\end{equation}
where $\tilde{s}_\mathrm{max}$ represents the maximum spacing error. 
Recall that the spacing error of the CAVs is contained in the system output~\eqref{Eq:SystemOutput}, whose future sequence $y$ serves as a decision variable in behavior representation~\eqref{Eq:AdaptedDeePCAchievability}. Thus, we translate the constraints~\eqref{Eq:Constraint_MinimumSpacing} and~\eqref{Eq:Constraint_MaximumSpacing} on the spacing errors to the following constraint on future output sequence
\begin{equation} \label{Eq:SafetyConstraint}
	\tilde{s}_\mathrm{min} \leq I_{N} \otimes \begin{bmatrix}
	\mathbb{0}_{m \times n} & I_m
	\end{bmatrix} y \leq \tilde{s}_\mathrm{max}.
\end{equation}

Finally, the control input of each CAV is constrained considering the vehicular actuation limit, given as follows
\begin{equation} \label{Eq:AccelerationConstraint}
		a_\mathrm{min} \leq u \leq a_\mathrm{max},
\end{equation} 
where $	a_\mathrm{min}$ and $a_\mathrm{max}$ denote the minimum and the maximum acceleration, respectively. 

\subsection{Formulation of \method{DeeP-LCC}}

We are now ready to present the following optimization problem to obtain the optimal control input of the CAVs
\begin{equation} \label{Eq:AdaptedDeePC}
\begin{aligned}
\min_{g,u,y} \quad &J(y,u)\\
\st \quad &\eqref{Eq:AdaptedDeePCAchievability},\eqref{Eq:FutureExternalInput},\eqref{Eq:SafetyConstraint},\eqref{Eq:AccelerationConstraint}.
\end{aligned}
\end{equation}
Note that unlike $u$ and $y$, the future velocity error sequence $\epsilon$ of the head vehicle, \ie, the external input of the mixed traffic system, is not a decision variable in~\eqref{Eq:AdaptedDeePC}; instead, it is fixed as a constant value, as shown in~\eqref{Eq:FutureExternalInput}. 

Further, it is worth noting that the non-parametric behavior representation shown in Proposition~\ref{Proposition:DeePLCC_FundamentalLemma} is valid for deterministic LTI mixed traffic systems.
In practice, the car-following behavior of HDVs is nonlinear, as discussed in Section~\ref{Sec:NonlinearTraffic}, and also has certain uncertainties, leading to a nonlinear and non-deterministic mixed traffic system. Practical traffic data collected from such a nonlinear system is also noise-corrupted, and thus the equality constraint~\eqref{Eq:AdaptedDeePCAchievability} becomes inconsistent, \ie, the subspace spanned by the columns of the data Hankel matrices fails to coincide with the subspace of all valid trajectories of the underlying system. 

 Motivated by the regulated version of DeePC~\cite{coulson2019data}, we introduce a slack variable $\sigma_y \in \mathbb{R}^{(n+m)T_\mathrm{ini}}$ for the system past output to ensure the feasibility of the equality constraint, and then solve the following regularized optimization problem 
  \begin{equation} \label{Eq:AdaptedDeePCforNonlinearSystem}
 \begin{aligned}
 \min_{g,u,y,\sigma_y} \quad &J(y,u)+\lambda_g \left\|g\right\|_2^2+\lambda_y \left\|\sigma_y\right\|_2^2\\
 \st \quad & \begin{bmatrix}
 U_\mathrm{p} \\ E_\mathrm{p}\\Y_\mathrm{p} \\ U_\mathrm{f} \\ E_\mathrm{f}\\ Y_\mathrm{f}
 \end{bmatrix}g=
 \begin{bmatrix}
 u_\mathrm{ini} \\ \epsilon_\mathrm{ini}\\ y_\mathrm{ini} \\ u \\\epsilon \\ y
 \end{bmatrix}+\begin{bmatrix}
 0\\0\\ \sigma_y \\0 \\0 \\0
 \end{bmatrix},\\ &\eqref{Eq:FutureExternalInput},\eqref{Eq:SafetyConstraint},\eqref{Eq:AccelerationConstraint}.
 \end{aligned}
 \end{equation}
 
{ This formulation \eqref{Eq:AdaptedDeePCforNonlinearSystem} is applicable to nonlinear and non-deterministic mixed traffic systems. 
In~\eqref{Eq:AdaptedDeePCforNonlinearSystem}, the slack variable $\sigma_y$ is penalized with a weighted two-norm penalty function, and the weight coefficient $\lambda_y>0$ can be chosen sufficiently large such that $\sigma_y \neq 0$ only if the equality constraint is infeasible. In addition, a two-norm penalty on $g$ with a weight coefficient $\lambda_g>0$ is also incorporated. Intuitively, the regularization term $\lambda_g \left\|g\right\|_2^2$ reduces the ``complexity" of the data-centric behavior representation and avoids overfitting, while the term $ \lambda_y \left\|\sigma_y\right\|_2^2$ improves the prediction accuracy whilst guaranteeing the representation feasibility. The introduction of} {   the practical constraints~\eqref{Eq:SafetyConstraint}, \eqref{Eq:AccelerationConstraint} provides safety guarantees for the CAVs when they are feasible. The notion of recursive feasibility plays a critical role for safety guarantees. We refer the interested readers to a recent result~\cite[Proposition 1]{berberich2020data} on recursive feasibility of the standard DeePC under an upper-level bounded condition on the slack variable $\sigma_y$ and a terminal constraint of stabilizing the system at equilibrium within the predictive horizon $N$. Due to the page limit, we leave the recursive feasibility of \method{DeeP-LCC} for future work.} 

 As shown in Fig.~\ref{Fig:SystemSchematic}, our proposed \method{DeeP-LCC} mainly consists of two parts: 
\begin{enumerate}
    \item \textit{offline data collection}, which records measurable input/output traffic data and constructs data Hankel matrices;
    \item \textit{online predictive control}, which relies on data-centric representation of system behavior for future trajectory prediction.
\end{enumerate}

In particular, at each time step during online predictive control, we solve the final \method{DeeP-LCC} formulation~\eqref{Eq:AdaptedDeePCforNonlinearSystem} in a receding horizon manner. 
For implementation, the optimization problem~\eqref{Eq:AdaptedDeePCforNonlinearSystem} is solved in a receding horizon manner. Algorithm~\ref{Alg:DeeP-LCC} lists the procedure of \method{DeeP-LCC}. We note that problem~\eqref{Eq:AdaptedDeePCforNonlinearSystem} amounts to solve a quadratic program, for which very efficient and reliable solvers exist.

\begin{algorithm}[t]
	\caption{\method{DeeP-LCC} for mixed traffic control}
	\label{Alg:DeeP-LCC}
	\begin{algorithmic}[1]
		\Require
		Pre-collected traffic data $(u^{\mathrm{d}},\epsilon^{\mathrm{d}},y^{\mathrm{d}})$, initial time $t_0$, terminal time $t_f$;
		\State Construct data Hankel matrices $U_\mathrm{p} , U_\mathrm{f}, E_\mathrm{p} , E_\mathrm{f}, Y_\mathrm{p} , Y_\mathrm{f}$;
		\State Initialize past traffic data $(u_{\mathrm{ini}},\epsilon_{\mathrm{ini}},y_{\mathrm{ini}})$ before the initial time $t_0$;
		\While{$t_0 \leq t \leq t_f$}
		\State Solve~\eqref{Eq:AdaptedDeePCforNonlinearSystem} for optimal predicted input $u^*=\col(u^*(t),u^*(t+1),\ldots,u^*(t+N-1))$;
		\State Apply the input $u(t) \leftarrow u^*(t)$ to the CAVs;
		\State $t \leftarrow t+1$ and update past traffic data $(u_{\mathrm{ini}},\epsilon_{\mathrm{ini}},y_{\mathrm{ini}})$;
		\EndWhile
	\end{algorithmic}
\end{algorithm}

{ 
\begin{remark}[Regularization]
The regularization approach in~\eqref{Eq:AdaptedDeePCforNonlinearSystem} is common in the recent work on employing Willems' fundamental and DeePC for nonlinear and stochastic control~\cite{coulson2019regularized,berberich2022linear,berberich2020data,elokda2021data,carlet2020data,huang2021decentralized}.  From a theoretic perspective, it has been revealed in~\cite{huang2021decentralized,coulson2019regularized} that the regulation on $g$ coincides with distributional robustness. Some closed-loop properties, such as recursive feasibility and exponential stability, have also been rigorously proved in~\cite{berberich2022linear,berberich2020data} for nonlinear and stochastic systems by imposing terminal constraints and typical auxiliary assumptions (\eg, linear independence constraint qualification). In addition, the effectiveness of the regularization has been demonstrated in multiple empirical studies on practical nonlinear systems with noisy measurements, including quadcopter systems~\cite{elokda2021data}, power grids~\cite{huang2021decentralized}, and electric motor drives~\cite{carlet2020data}. Motivated by the aforementioned research, we introduce this regularization into \method{DeeP-LCC} for the nonlinear and non-deterministic traffic systems. Note that unlike previous work~\cite{huang2021decentralized,coulson2019regularized,berberich2022linear,berberich2020data}, our formulation~\eqref{Eq:AdaptedDeePCforNonlinearSystem} has an external disturbance input signal $\epsilon$, and we leave its theoretical investigation for future research. Indeed, it is observed from our nonlinear simulations in Section~\ref{Sec:6} and our follow-up real-world miniature experiments in~\cite{wang2022implementation} that the proposed regularized formulation~\eqref{Eq:AdaptedDeePCforNonlinearSystem} achieves effective wave-dampening performance for CAVs in practical mixed traffic systems. 
\markend
\end{remark}}

\begin{remark}[External input]
Compared to standard DeePC, we introduce the external input signal and utilize~\eqref{Eq:FutureExternalInput} to predict its future value. To address the unknown future external input, another approach is to assume a bounded future velocity error of the head vehicle. This idea is similar to robust DeePC against unknown external disturbances; see, \eg,~\cite{huang2021decentralized,lian2021adaptive}. 
It is interesting to further design robust DeePC for mixed traffic when the head vehicle is oscillating around an equilibrium velocity, but this is beyond the scope of this work. In the next section, our traffic simulations reveal that by  assuming~\eqref{Eq:FutureExternalInput} and updating equilibrium based on historical velocity data of the head vehicle, the proposed \method{DeeP-LCC} has already shown excellent performance in improving traffic performance.
\markend
\end{remark}

\begin{remark}[Computational complexity]
\label{remark:complexity}
 
For mixed traffic control, both MPC and \method{DeeP-LCC} can be formulated into a quadratic program for numerical computation. In the \method{DeeP-LCC} formulation~\eqref{Eq:AdaptedDeePCforNonlinearSystem}, one could use $g\in \mathbb{R}^{T-T_\mathrm{ini}-N+1}$ as the main decision variable, with an equality constraint given by $
    U_{\mathrm{p}} g = 
    u_{\mathrm{ini}} \in \mathbb{R}^{T_\mathrm{ini}m}
$ (here the influence of the external input is neglected without loss of generality). As revealed in~\eqref{Eq:DataLength}, the pre-collected data length $T$ is lower bounded by $(m+1)(T_{\mathrm{ini}}+N+2n)-1$, and thus \method{DeeP-LCC} has at least $2mn+2n+Nm$ free decision variables. For MPC, its optimization size is captured by the future control sequence $u \in \mathbb{R}^{Nm}$. Therefore, the online optimization size of \method{DeeP-LCC} is slightly larger than that of MPC with $2mn+2n$ additional decision variables, but it is observed in Section~\ref{Sec:6} that the computation time of \method{DeeP-LCC} is acceptable for small-scale simulations (about $28.07\,\mathrm{ms}$). Meanwhile, the simplicity of \method{DeeP-LCC} is worth noting: it directly utilizes a single trajectory for online predictive control based on one integrated optimization formulation~\eqref{Eq:AdaptedDeePCforNonlinearSystem}. Particularly, it requires no prior knowledge of the system model, and circumvents an offline model identification step and an online initial state estimation step, which are necessary steps in standard output-feedback MPC. Still, it is an important future direction to improve the computational efficiency of \method{DeeP-LCC} for large-scale mixed traffic flow. We refer the interested readers to~\cite{wang2022distributed} for a recent potential approach by distributed optimization.
\markend
\end{remark}

\section{Traffic Simulations}
\label{Sec:6}

This section presents three nonlinear and non-deterministic traffic simulations to validate the performance of \method{DeeP-LCC} in mixed traffic. The nonlinear OVM model~\eqref{Eq:OVMmodel} is utilized to depict the dynamics of HDVs. { A noise signal with the uniform distribution of $\mathbb{U}[-0.1,0.1]\,\mathrm{m/s^2}$ is added  to acceleration dynamics model~\eqref{Eq:OVMmodel} of each HDV in our simulations.}\footnote{
The algorithm and simulation scripts are available at \url{https://github.com/soc-ucsd/DeeP-LCC}.} 

For the mixed traffic system in Fig.~\ref{Fig:SystemSchematic}, we consider eight vehicles behind the head vehicle, among which there exist two CAVs and six HDVs, \ie, $n=8$, $m=2$ (this corresponds to a CAV penetration rate of $25\%$). The two CAVs are located at the third and the sixth vehicles respectively, \ie, $S=\{3,6\}$. 
The parameter setup for \method{DeeP-LCC} is as follows. 

\begin{itemize}
    \item \emph{Offline data collection:} the length for the pre-collected trajectory is chosen as $T=800$ with a sampling interval $\Delta t = 0.05 \,\mathrm{s}$. { When collecting trajectories, we consider an equilibrium traffic velocity of $15\,\mathrm{m/s}$, and the pre-collected data sets from this equilibrium are used for all the following experiments. For the system inputs, we utilize the OVM model~\eqref{Eq:OVMmodel} as a pre-designed controller for the CAVs with random noise perturbations, and assume a random slight perturbation on the head vehicle's velocity. Given a sufficiently long trajectory, this design naturally satisfies the persistent excitation requirement in Assumption~\ref{Assumption:PersistentExcitation} and is also applicable to practical traffic flow. More details and an illustration of a pre-collected trajectory 
    can be found in Appendix~\ref{Appendix:DataCollection}.}
    \item \emph{Online control procedure:} the time horizons for the future signal sequence and past signal sequence are set to $N=50$, $T_{\mathrm{ini}}=20$, respectively. In the cost function~\eqref{Eq:CostDefinition}, the weight coefficients are set to $w_v=1,w_s=0.5,w_u=0.1$; { for constraints, the boundaries for the spacing of the CAVs are set to $s_{\max} = 40\,\mathrm{m}, s_{\min} = 5 \,\mathrm{m}$, and the limit for the acceleration of the CAVs are set to $a_{\max} = 2 \,\mathrm{m/s^2}, a_{\min} = -5\,\mathrm{m/s^2}$} (this limit also holds for all the HDVs via saturation). In the regulated formulation~\eqref{Eq:AdaptedDeePCforNonlinearSystem},
the parameters are set to $\lambda_g=10,\lambda_y=10000$. 
\end{itemize}

\subsection{Performance Validation around an Equilibrium State}
\label{Sec:Simulation1}

\begin{figure}[t]
	\vspace{1mm}
	\centering
	\subfigure[All HDVs]
	{\includegraphics[width=4cm]{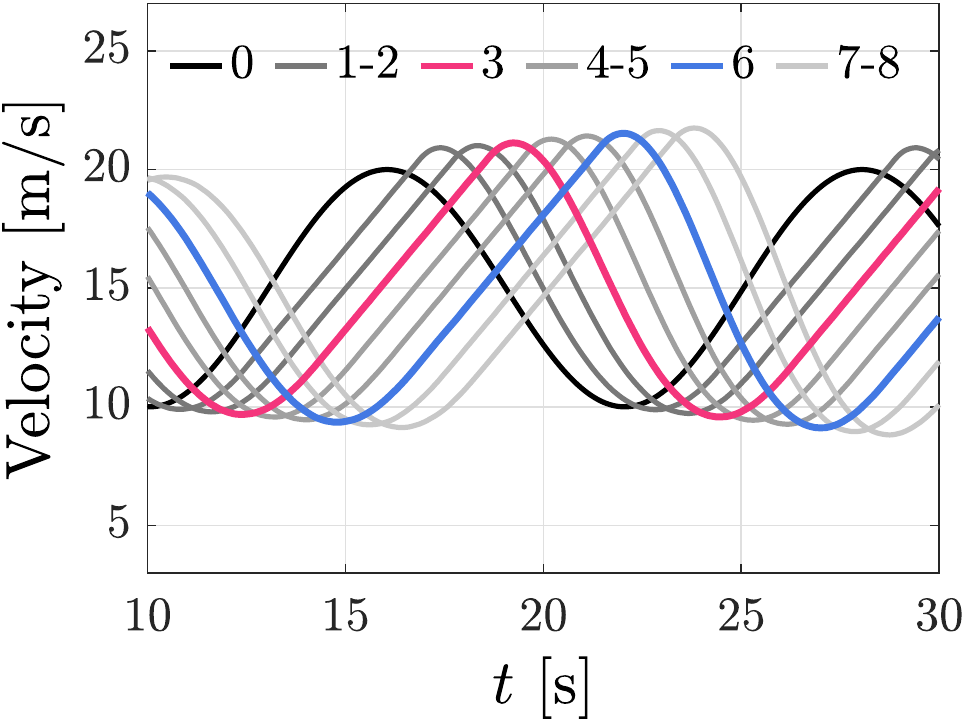}
	\label{Fig:SinusoidPerturbation_HDVs}}\\
 \subfigure[MPC]
	{\includegraphics[width=4cm]{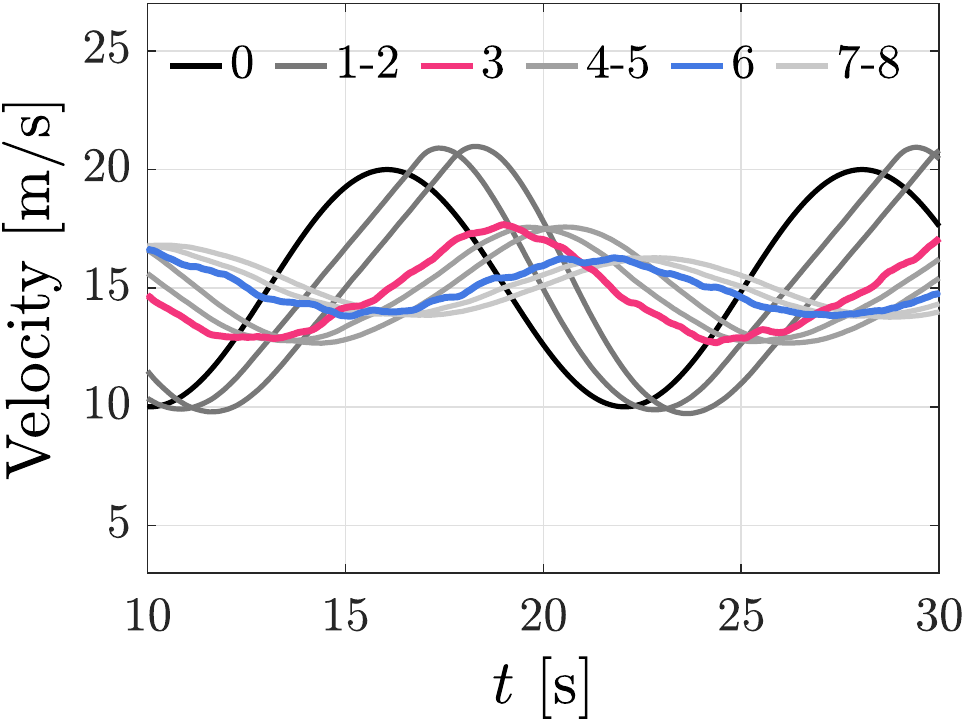}
	\label{Fig:SinusoidPerturbation_MPC}}
	\subfigure[\method{DeeP-LCC}]
	{\includegraphics[width=4cm]{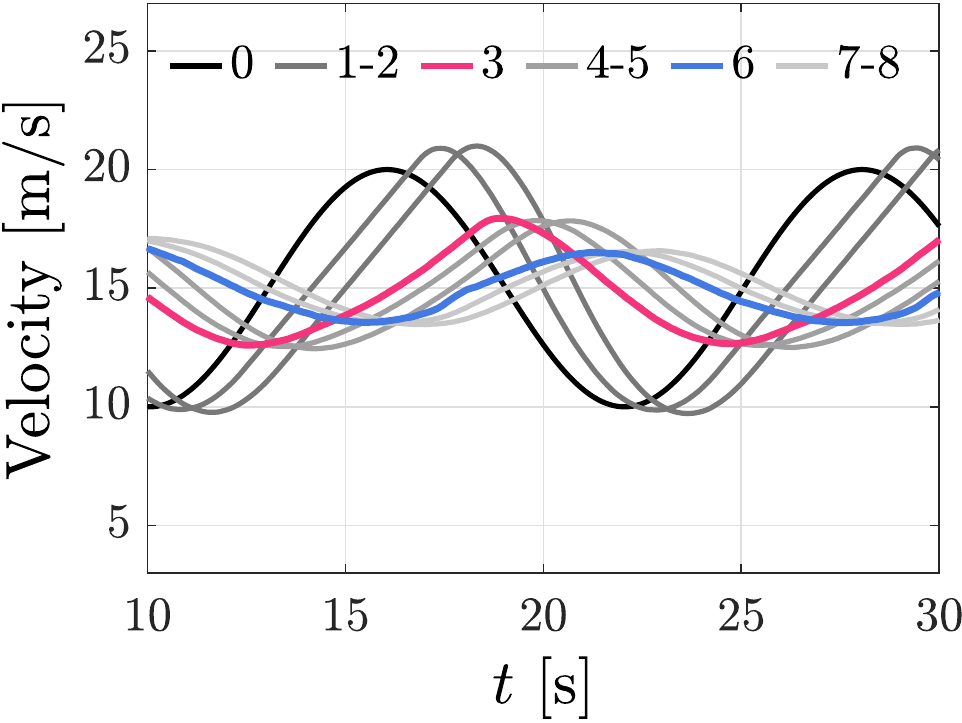}
	\label{Fig:SinusoidPerturbation_DeePC}}
	\vspace{-1mm}
	\caption{{ Velocity profiles in Experiment A, where a sinusoidal perturbation is imposed on the head vehicle. The black profile represents the head vehicle (vehicle $0$, and the gray profile represents the HDVs with different darkness denoting different vehicle indices. The red profile (vehicle $3$) and the blue profile (vehicle $6$) represent the first and the second CAV, respectively.} (a) All the vehicles are HDVs. (b) The CAVs utilize the MPC controller. (c) The CAVs utilize the \method{DeeP-LCC} controller.}
	\label{Fig:SinusoidPerturbation_VelocityProfile}
\end{figure}

{ Motivated by~\cite{jin2017optimal
,di2019cooperative,huang2020learning}, our first experiment (Experiment A) simulates 
a traffic wave scenario, where vehicles accelerate and decelerate periodically, by imposing a sinusoidal perturbation on the head vehicle around the equilibrium velocity of $15\, \mathrm{m/s}$ (see the black profile in Fig.~\ref{Fig:SinusoidPerturbation_VelocityProfile} for the velocity trajectory of the head vehicle), and investigates the performance of CAVs in dampening traffic waves.} Particularly, we aim to compare the performance of the proposed \method{DeeP-LCC}  with the standard output-feedback MPC~\eqref{Eq:MPC} based on an accurate mixed traffic system model~\eqref{Eq:DT_TrafficModel}. The dynamical model for all the HDVs is set to follow the nominal parameter values~\cite{jin2017optimal,zheng2020smoothing,wang2020controllability}: $\alpha=0.6, \beta=0.9, v_{\max }=30, s_{\mathrm{st}}=5, s_{\mathrm{go}}=35, v^* = 15$. { The MPC controller is designed using the accurate linearized model~\eqref{Eq:DT_TrafficModel} around the same equilibrium velocity of $15\,\mathrm{m/s}$ as that in offline data collection}, while \method{DeeP-LCC} 
is designed according to the procedures in Section~\ref{Sec:5}. The other parameters, \eg, the coefficients in cost function and past/future time horizon, remain the same between MPC and \method{DeeP-LCC}.

When all the vehicles are HDVs, it is observed in Fig.~\ref{Fig:SinusoidPerturbation_HDVs} that the amplitude of such perturbation is amplified along the propagation. This perturbation amplification greatly increases fuel consumption and collision risk in mixed traffic. By contrast, with two CAVs existing in traffic flow and~employing either MPC or \texttt{DeeP-LCC}, the amplitude of the perturbation is clearly attenuated, as shown in Fig.~\ref{Fig:SinusoidPerturbation_MPC} and Fig.~\ref{Fig:SinusoidPerturbation_DeePC}, respectively. This demonstrates the capabilities of CAVs in dissipating undesired disturbances and stabilizing traffic flow using either MPC or  \method{DeeP-LCC}.

\begin{figure}[t]
	\vspace{1mm}
	\centering
	\includegraphics[scale=0.45]{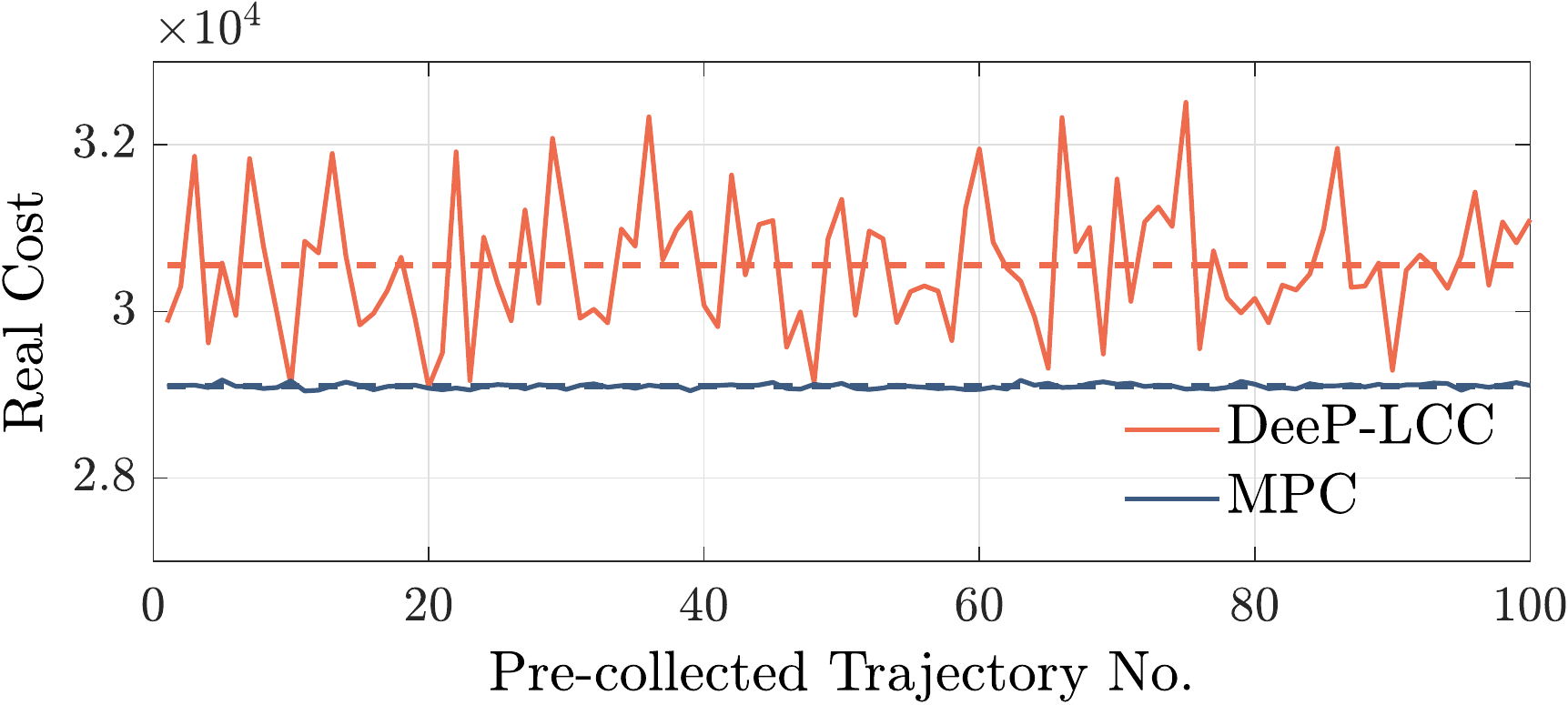}
	\vspace{-1mm}
	\caption{Comparison of real cost between  \method{DeeP-LCC} and MPC in 100 experiments in Experiment A. The dashed line represents the average real cost of each controller, { which is $2.91\times 10^4$ for MPC and $3.05\times 10^4$ for \method{DeeP-LCC} with a standard deviation of $0.003\times 10^4$ and $0.077\times 10^4$, respectively}.}
	\label{Fig:SinusoidPerturbation_RealCost}
\end{figure}

We note that Fig.~\ref{Fig:SinusoidPerturbation_DeePC} demonstrates the performance of  \method{DeeP-LCC} using one single pre-collected trajectory. Different pre-collected trajectories might influence the performance of \method{DeeP-LCC}, as  \method{DeeP-LCC} directly relies on these data  to design the CAVs' control input. To see the influence, we collect $100$ trajectories of the same length $T=800$ to construct the data Hankel matrices~\eqref{Eq:AdaptedDeePCforNonlinearSystem} and carry out the same experiment. Fig.~\ref{Fig:SinusoidPerturbation_RealCost} shows the  cost value $J$ given by~\eqref{Eq:CostDefinition} at each simulation under  \method{DeeP-LCC} or MPC. { Recall that MPC utilizes the accurate linearized dynamics for control input design, and its performance can be regarded as the optimal benchmark for the nonlinear traffic control around the equilibrium state. In comparison, \method{DeeP-LCC} directly relies on the raw trajectory data, and the regularization in~\eqref{Eq:AdaptedDeePCforNonlinearSystem} might influence the optimality of the original cost $J(y,u)$. 
From our random experiments, we observe that \method{DeeP-LCC} achieves a mean real cost that is quite close to the benchmark (losing only $4.8\%$ optimality) for the noise-corrupted nonlinear traffic system \emph{without requiring any knowledge of the underlying system}. These random experimental results validate the comparable wave-dampening performance of \method{DeeP-LCC} with respect to MPC based on accurate dynamics. 
This observation is consistent with previous studies of DeePC on other nonlinear dynamical systems such as quadcopters~\cite{coulson2019data} or power grid~\cite{huang2021decentralized}.}


\begin{table}[t]
	\begin{center}
		\caption{Heterogeneous Parameter Setup for HDVs\\ in Experiments B and C}\label{Tb:HDVParameterSetup}
		\begin{threeparttable}
		\setlength{\tabcolsep}{7mm}{
		\begin{tabular}{cccc}
		\toprule
			& $\alpha$ & $\beta$ & $s_{\mathrm{go}}$ \\\hline
			HDV 1 & 0.45& 0.60& 38\\
			HDV 2 & 0.75& 0.95& 31\\
			HDV 3 & 0.70& 0.95& 33\\
			HDV 4 & 0.50& 0.75& 37\\
			HDV 5 & 0.40& 0.80& 39\\
			HDV 6 & 0.80& 1.00& 34\\
			Nominal Setup & 0.60& 0.90& 35 \\
			\bottomrule
		\end{tabular}}
		\begin{tablenotes}
		\footnotesize
		\item[1] The HDVs are indexed from front to end. For example, HDV~1 and HDV~2 are the two HDVs between the head vehicle and the first CAV.
		\item[2] The other parameters follow the nominal setup: $s_\mathrm{st} = 5, v_{\max}=30$.
		\end{tablenotes}
		\end{threeparttable}
	\end{center}
\end{table}

\begin{figure*}[t]
	\vspace{1mm}
	\centering
 \subfigure[All HDVs in Urban Scenarios]
	{
	\includegraphics[scale=0.46]{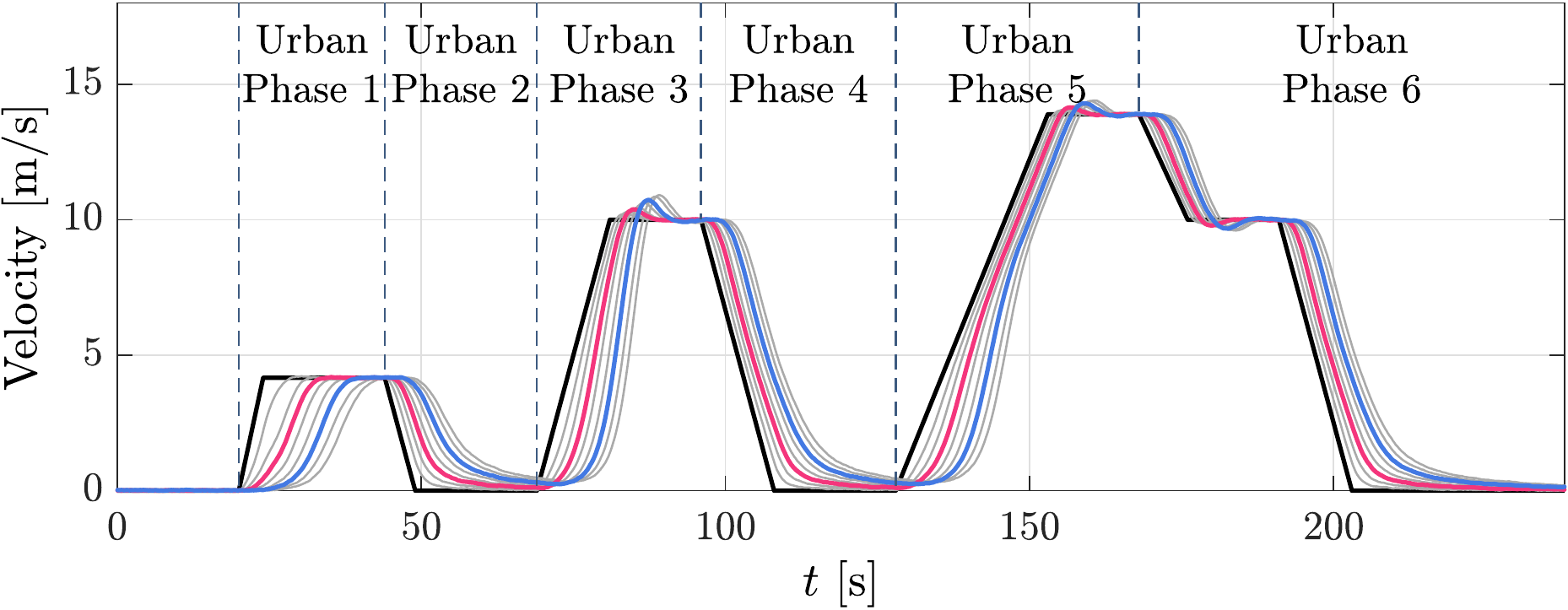}
 \label{Fig:NEDCSimulation_Urban_HDV}
	}
	\subfigure[All HDVs in Highway Scenarios]
	{
	\includegraphics[scale=0.46]{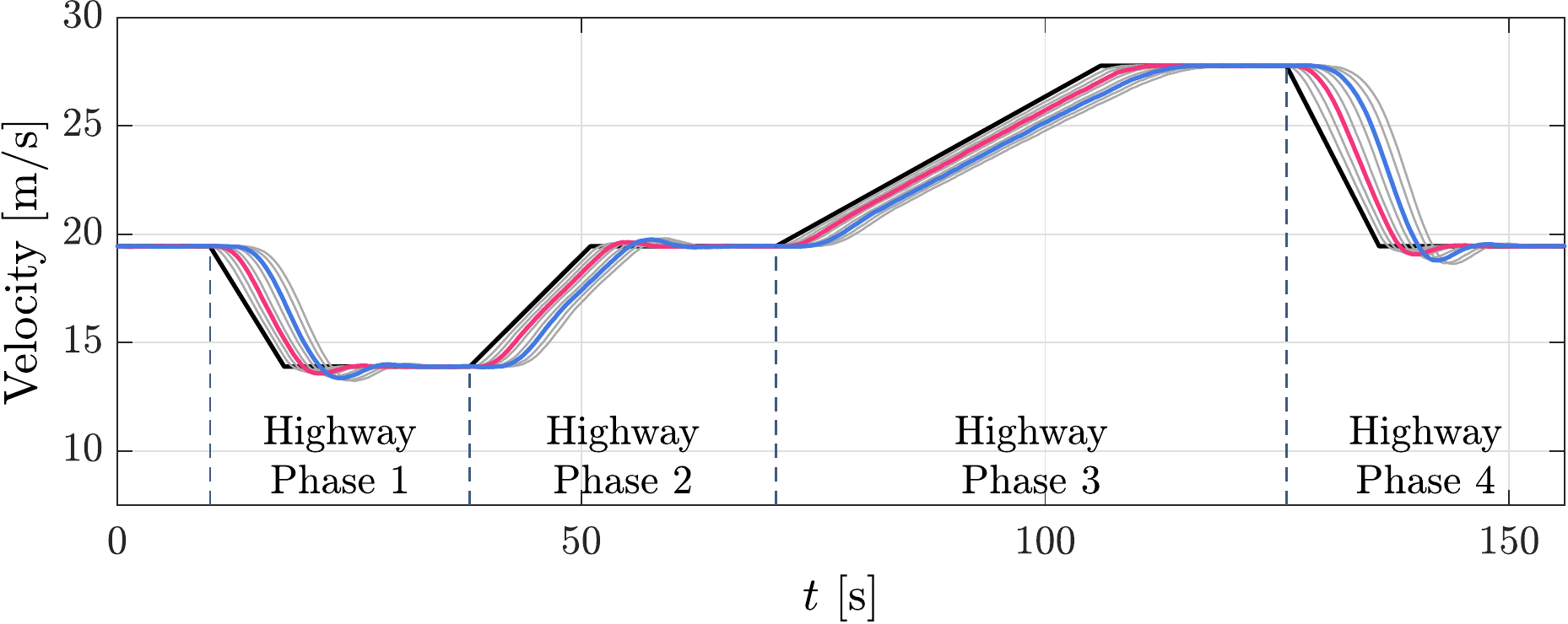}
	}
 \subfigure[\method{DeeP-LCC} in Urban Scenarios]
	{
	\includegraphics[scale=0.46]{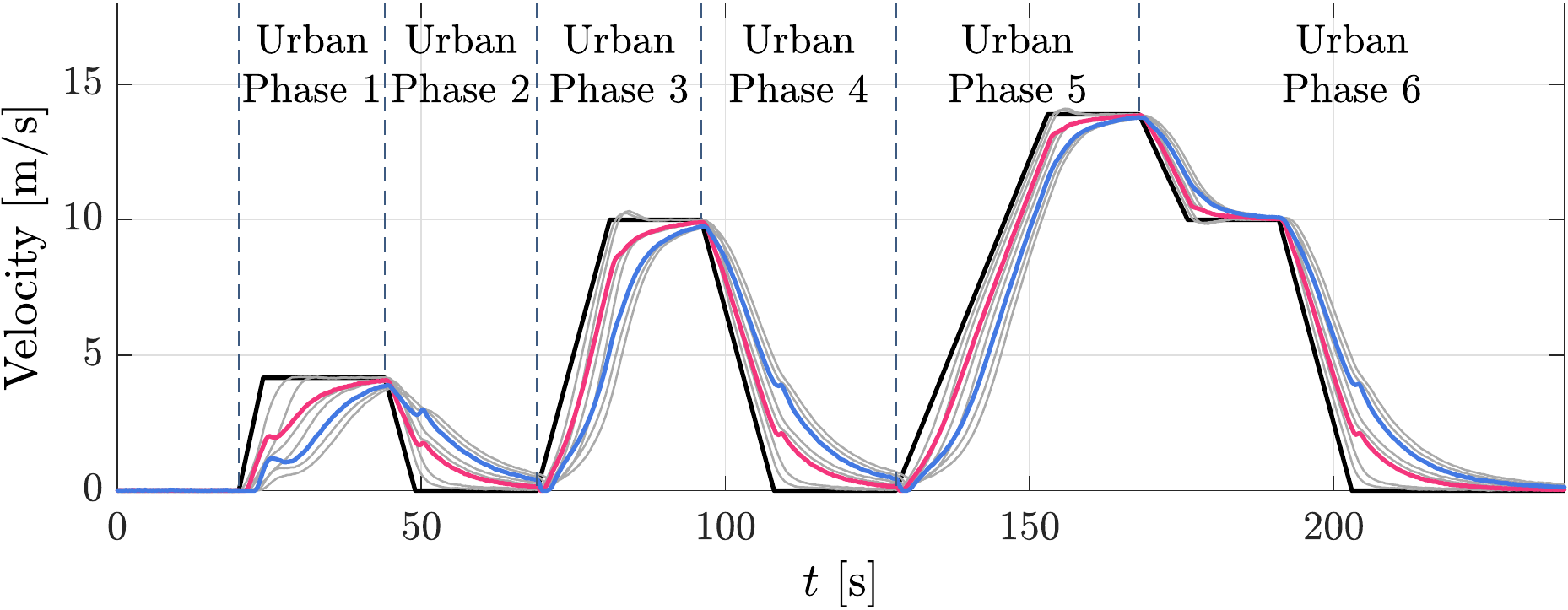}
  \label{Fig:NEDCSimulation_Urban_CAV}
	}
	\subfigure[\method{DeeP-LCC} in Highway Scenarios]
	{
	\includegraphics[scale=0.46]{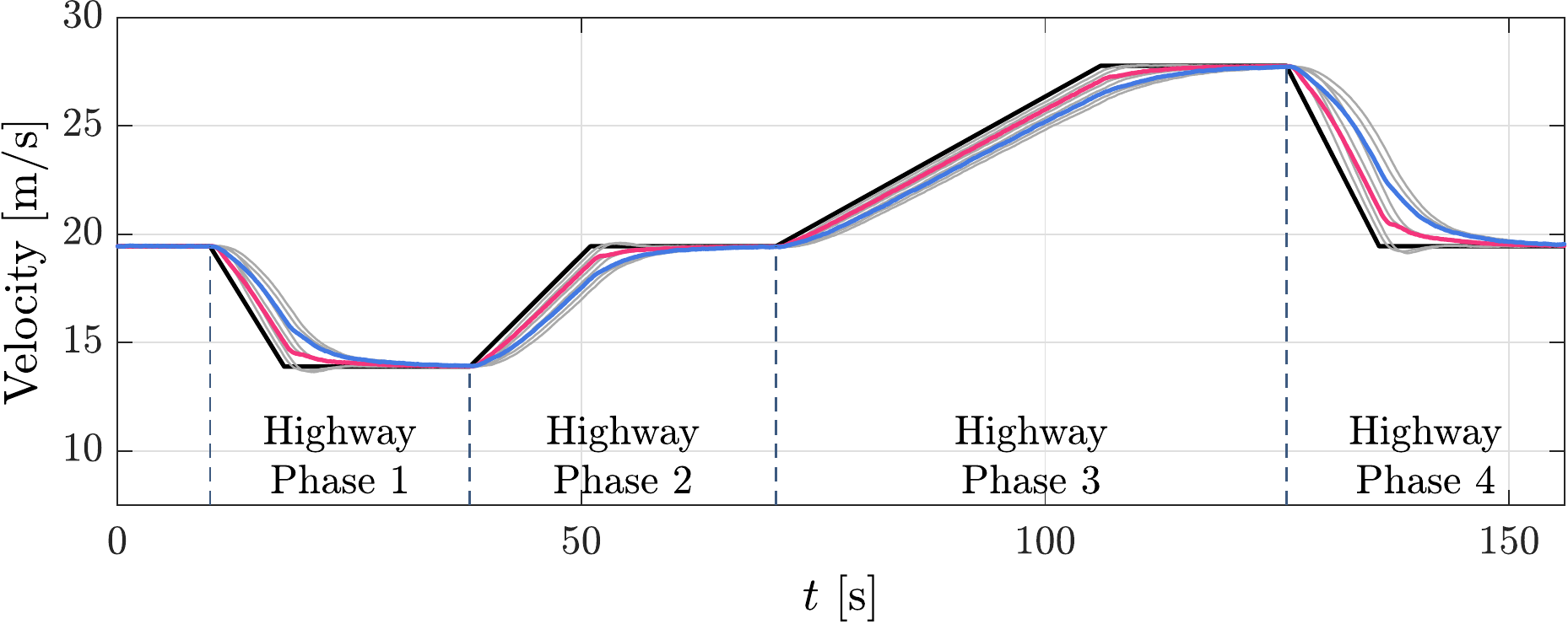}
	}
	\vspace{-1mm}
	\caption{ Velocity profiles in Experiment B, which is designed motivated by NEDC. (a)(b) denote the case where all the vehicles are HDVs in the urban and highway scenarios repsectively, while (c)(d) denote the case where there are two CAVs utilizing \method{DeeP-LCC}. The color of each profile has the same meaning as that in Fig.~\ref{Fig:SinusoidPerturbation_VelocityProfile}.}
	\label{Fig:NEDCSimulation}
	\vspace{-4mm}
\end{figure*}

\subsection{Traffic Improvement in Comprehensive Simulation}
\label{Sec:Simulation2}

In Experiment A, we consider a fixed traffic equilibrium state and a nominal parameter setup for all HDVs. { Here in Experiment B, we design both an urban driving trajectory and a highway driving trajectory for the head vehicle motivated by ECE-15 and Extra-Urban Driving Cycle (EUDC) from the New European Driving Cycle (NEDC)~\cite{dieselnet2013emission}, and validate the capability of \method{DeeP-LCC} in improving traffic performance with time-varying equilibrium states.} In addition, we assume a heterogeneous parameter setup around the nominal value for all the HDVs by utilizing the OVM model~\eqref{Eq:OVMmodel}; see Table~\ref{Tb:HDVParameterSetup}. The MPC controller still utilizes the nominal parameter setup to design the control input, while   \method{DeeP-LCC} relies on pre-collected trajectory data as usual. { Note that practical traffic flow might have different equilibrium states in different time periods. In \method{DeeP-LCC}, we design a simple strategy to estimate equilibrium velocity by calculating the mean velocity of the head vehicle during the past horizon $T_{\mathrm{ini}}$ (the same time horizon for past signal sequence in  \method{DeeP-LCC}).} Meanwhile, the equilibrium spacing for the CAVs is chosen according to~\eqref{Eq:EquilibriumEquation_OVM} using the OVM model with a nominal parameter setup; see Appendix~\ref{Appendix:DeePC} for more details.

To quantify traffic performance, we consider the fuel consumption and velocity errors for the vehicles indexed from $3$ to $8$,  since the first two HDVs cannot be influenced by the CAVs (recall that $n=8$ and $S = \{3,6\}$). Precisely, we utilize an instantaneous fuel consumption model in~\cite{bowyer1985guide}: the fuel consumption rate $f_i$ ($\mathrm{mL/s}$) of the $i$-th vehicle  is calculated as
	\begin{equation*}
	f_i = \begin{cases}
	0.444+0.090 R_i v_i + [0.054 a_i^2 v_i] _{a_i>0},& \text{if}\; R_i>0,\\
	0.444, & \text{if} \; R_i \le 0,
	\end{cases}
	\end{equation*}
where $R_i = 0.333+0.00108 v_i^2 + 1.200 a_i$ with $a_i$ denoting the acceleration of vehicle $i$. To quantify velocity errors, we use an index of mean squared velocity error (MSVE) given by 
\begin{equation*}
\mathrm{MSVE} = \frac{\Delta t}{n(t_f-t_0)}\sum_{t=t_0}^{t_f}\sum_{i=1}^n(v_{i}(t)-v_0(t))^2,
\end{equation*}
where $t_0,t_f$ denote the begin and end time of the simulation respectively. This MSVE index depicts the tracking performance  towards the velocity of the head vehicle and measures traffic smoothness. 

\begin{figure}[t]
	\vspace{1mm}
	\centering
	\subfigure[Fuel Consumption Reduction Rate]
	{\includegraphics[width=8cm]{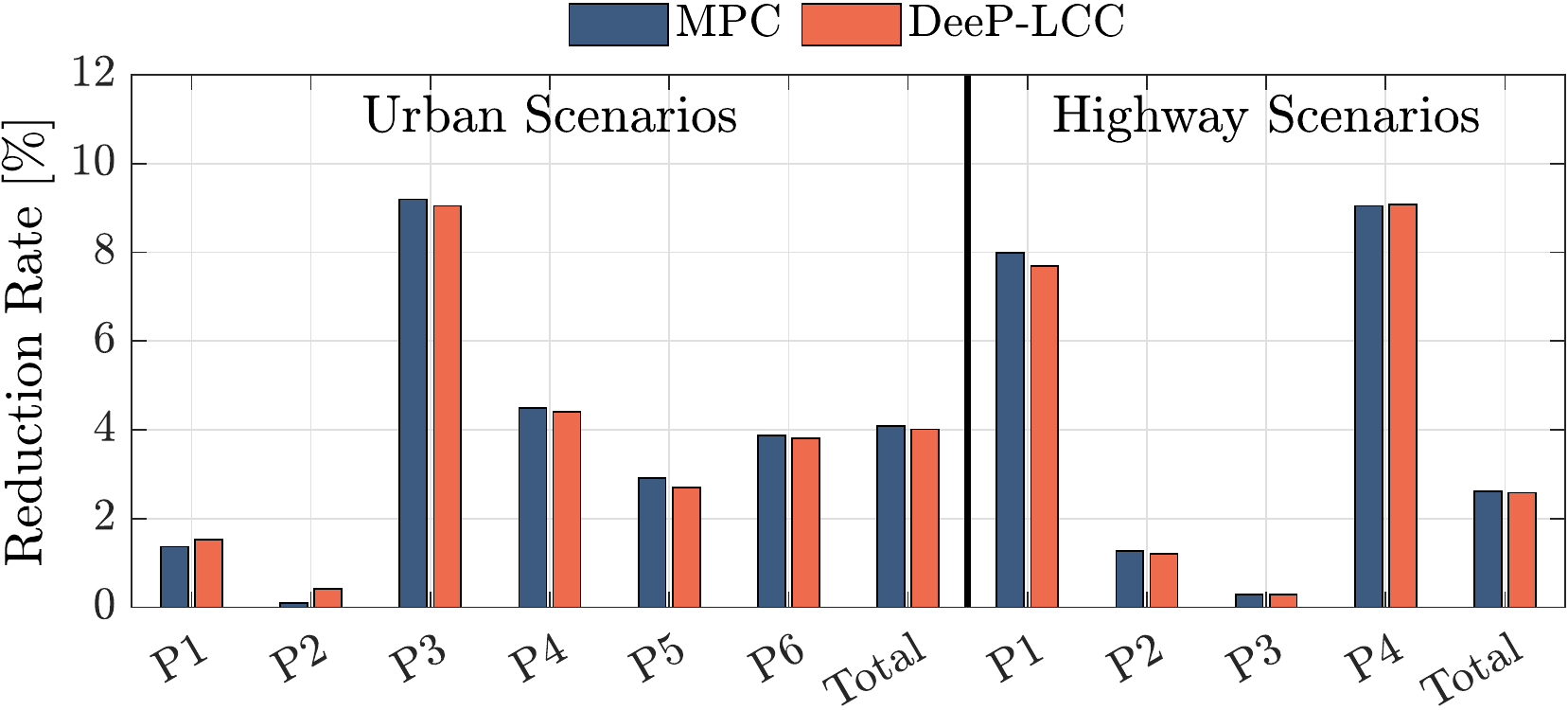}
	\label{Fig:NEDC_Fuel}}
	\subfigure[MSVE Reduction Rate]
	{\includegraphics[width=8cm]{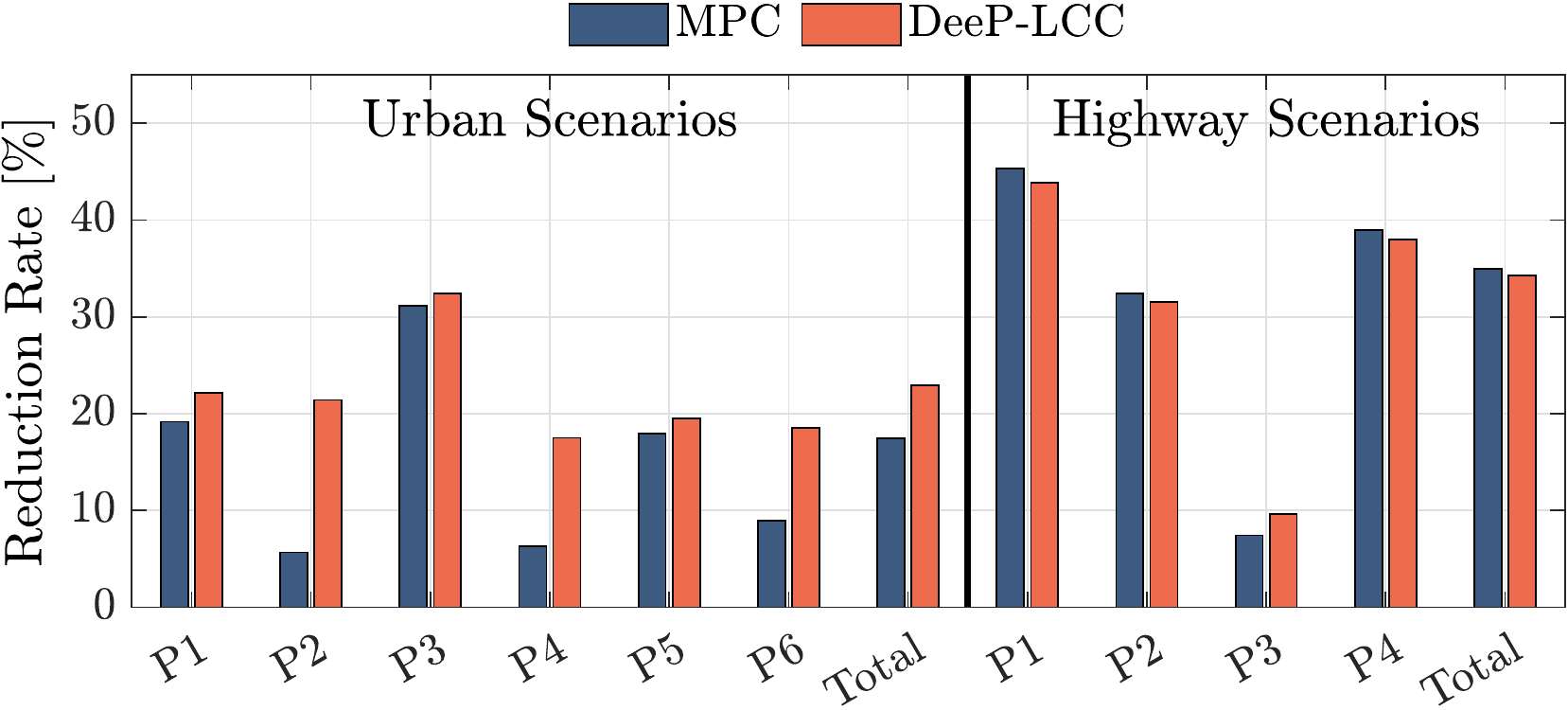}
	\label{Fig:NEDC_Error}}
	\vspace{-1mm}
	\caption{ Performance improvement of MPC and \method{DeeP-LCC} compared with the case where all the vehicles are HDVs in the comprehensive experiments. In the horizontal axis of each panel, P\# is abbreviated for Phase \#, denoting the phase number shown in Fig.~\ref{Fig:NEDCSimulation}.}
	\label{Fig:NEDC_Statistics}
	\vspace{-2mm}
\end{figure}

The results of velocity trajectories of each vehicle are shown in Fig.~\ref{Fig:NEDCSimulation}. {  Compared to the case with all HDVs,  \method{DeeP-LCC} allows the CAVs to rapidly track the trajectory of the head vehicle without overshoot, and thus mitigates velocity perturbations and smooths the mixed traffic flow in both urban and highway scenarios.  In addition, we observe that the improved traffic behavior under \method{DeeP-LCC} is close to that under MPC. 
By dividing the urban and highway driving cycles into different phases (see Fig.~\ref{Fig:NEDCSimulation}), we illustrate the reduction rate of fuel consumption and MSVE by MPC and \method{DeeP-LCC} with respect to the case with all HDVs in Fig.~\ref{Fig:NEDC_Statistics}. Both MPC and \method{DeeP-LCC} contribute to a significant improvement in fuel economy and traffic smoothness. In particular, \method{DeeP-LCC} saves up to $9.05\%$ fuel consumption during Phase 3 of urban driving scenarios and up to $43.82\%$ velocity error during Phase 1 of highway driving scenarios.}

{ Note that MPC utilizes the nominal model to design the control input, while \method{DeeP-LCC} relies on the trajectory data to directly predict the future system behavior. Thus, MPC  is not easily applicable in practice, since the nominal model for individual HDVs is generally unknown. By contrast, without explicitly identifying a parametric model, \method{DeeP-LCC} achieves a comparable performance with MPC using only pre-collected trajectory data, which are easier to acquire for the} { CAVs via V2V/V2I communications. In addition, although the} {  optimization complexity of \method{DeeP-LCC} is slightly higher than MPC (see Remark~\ref{remark:complexity}), its mean computation time during this experiment is $28.07\,\mathrm{ms}$ in a laptop computer equipped with Intel Core i7-11800H CPU and 32 G RAM. This computational cost is acceptable for real-time implementation in the underlying system scale ($8$ vehicles with $2$ CAVs).} 

{ 
\begin{remark}
Previous work on validating CAVs' wave-dampening performance mostly considers a similar simulation scenario to Experiments A and C (sinusoidal or brake perturbation); see, \eg,~\cite{jin2017optimal,di2019cooperative,wang2020controllability,huang2020learning}. In our work, we have introduced the driving cycle, which is indeed mostly used for measuring fuel consumption and emission of one single vehicle, to further demonstrate the performance of \method{DeeP-LCC} in various traffic scenarios. Additionally, note that in our data collection for \method{DeeP-LCC}, the traffic conditions around the fixed equilibrium velocity of $15\,\mathrm{m/s}$ are considered to capture the system behavior; see Appendix~\ref{Appendix:DataCollection} for illustration of the pre-collected trajectory. In the simulations, however, the equilibrium is time-varying, and we assume that the HDVs have a similar behavior around different equilibrium states in order to make the fundamental lemma directly applicable with the data collected from one single equilibrium. This assumption indeed may not hold, and thus the performance of \method{DeeP-LCC} might be compromised in this simulation. We provide further discussions and potential approaches to address time-varying equilibrium in Appendix~\ref{Appendix:DeePC}.
\markend
\end{remark}
}

\begin{figure}[t]
	\vspace{1mm}
	\centering
	\subfigure[]
	{\includegraphics[scale=0.42]{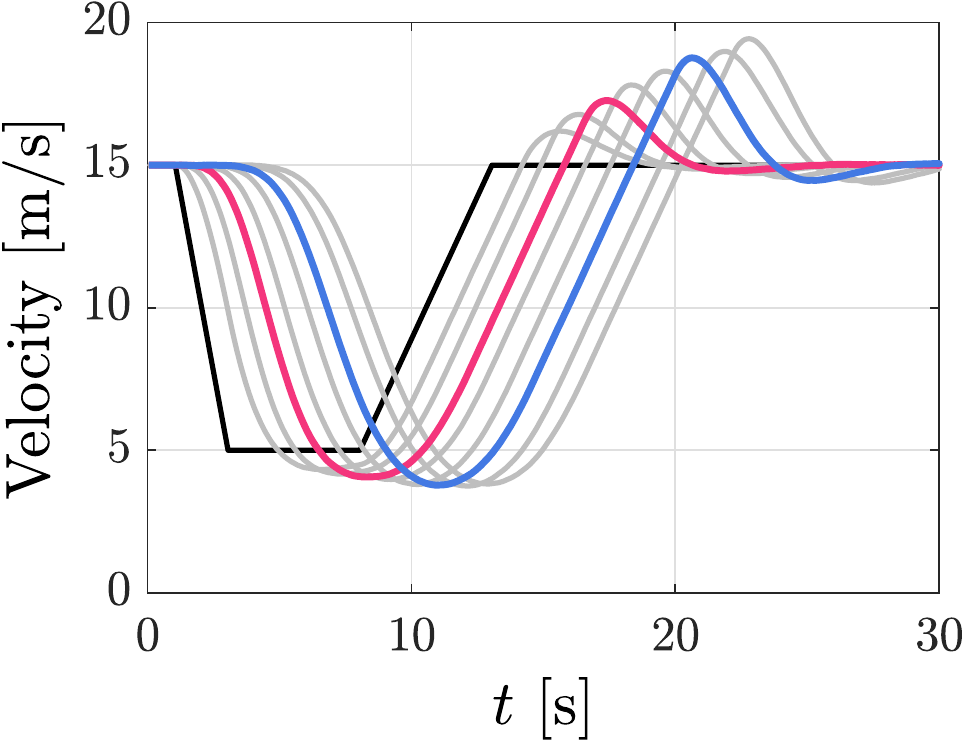}
	\label{Fig:BrakePerturbation_HDVs_Velocity}}
	\subfigure[]
	{\includegraphics[scale=0.42]{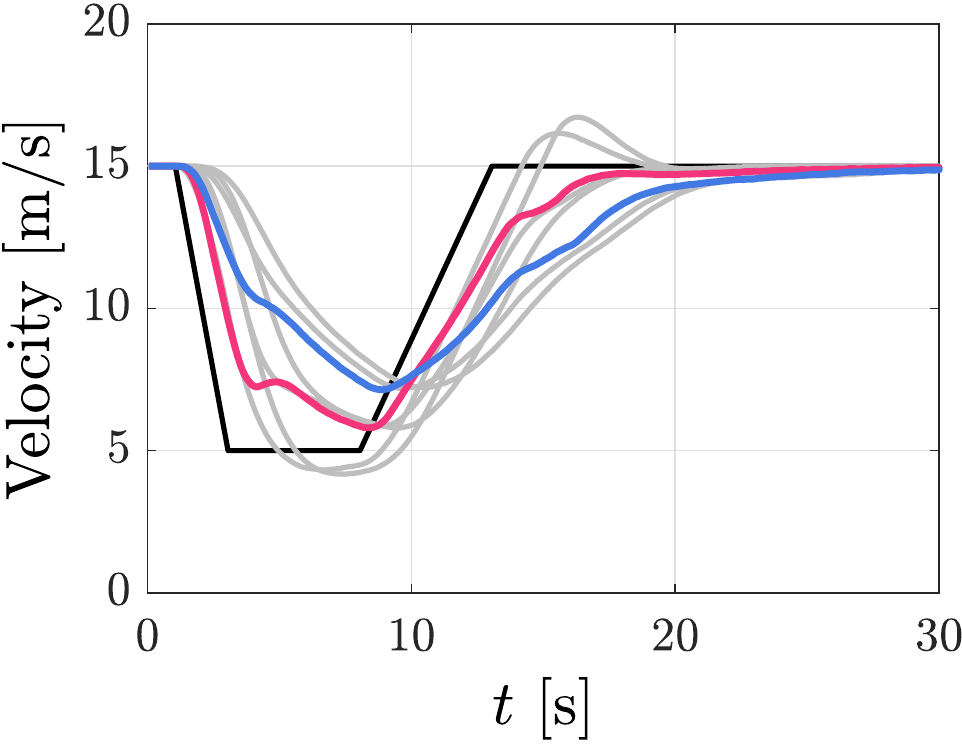}
	\label{Fig:BrakePerturbation_DeePC_Velocity}}
	\subfigure[]
	{\includegraphics[scale=0.42]{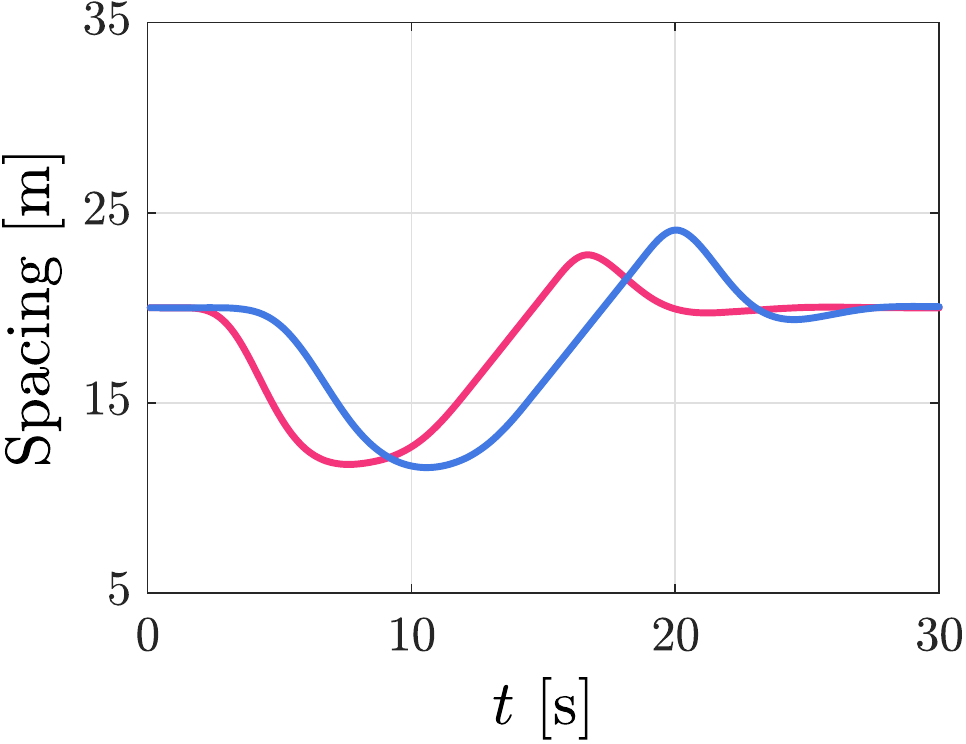}
	\label{Fig:BrakePerturbation_HDVs_Spacing}}
	\subfigure[]
	{\includegraphics[scale=0.42]{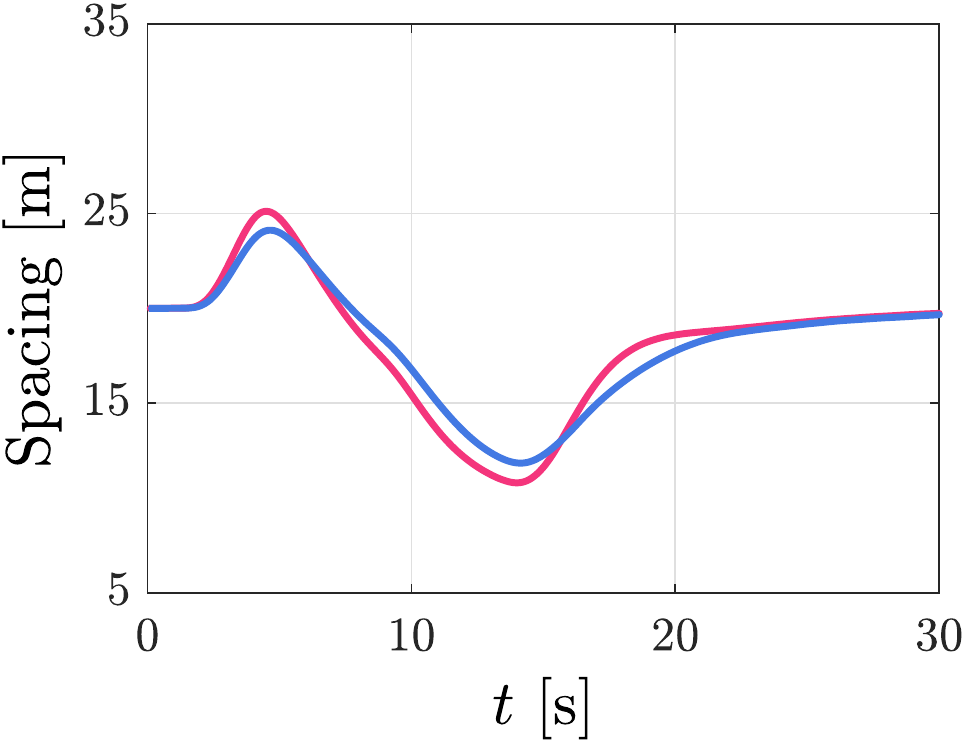}
	\label{Fig:BrakePerturbation_DeePC_Spacing}}
	\subfigure[]
	{\includegraphics[scale=0.42]{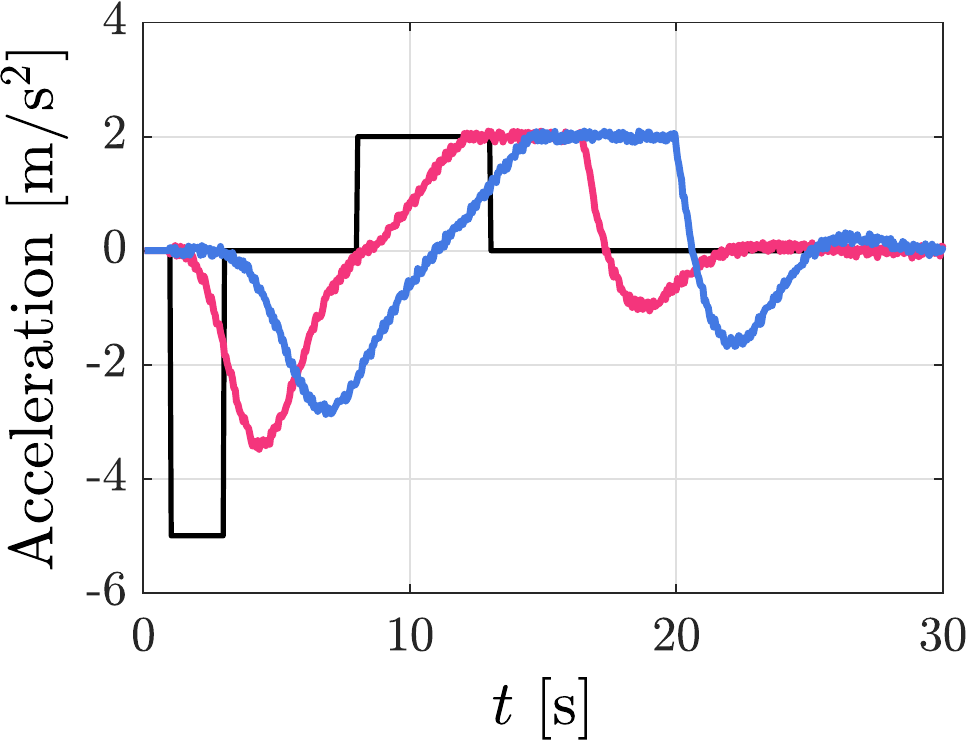}
	\label{Fig:BrakePerturbation_HDVs_Acceleration}}
	\subfigure[]
	{\includegraphics[scale=0.42]{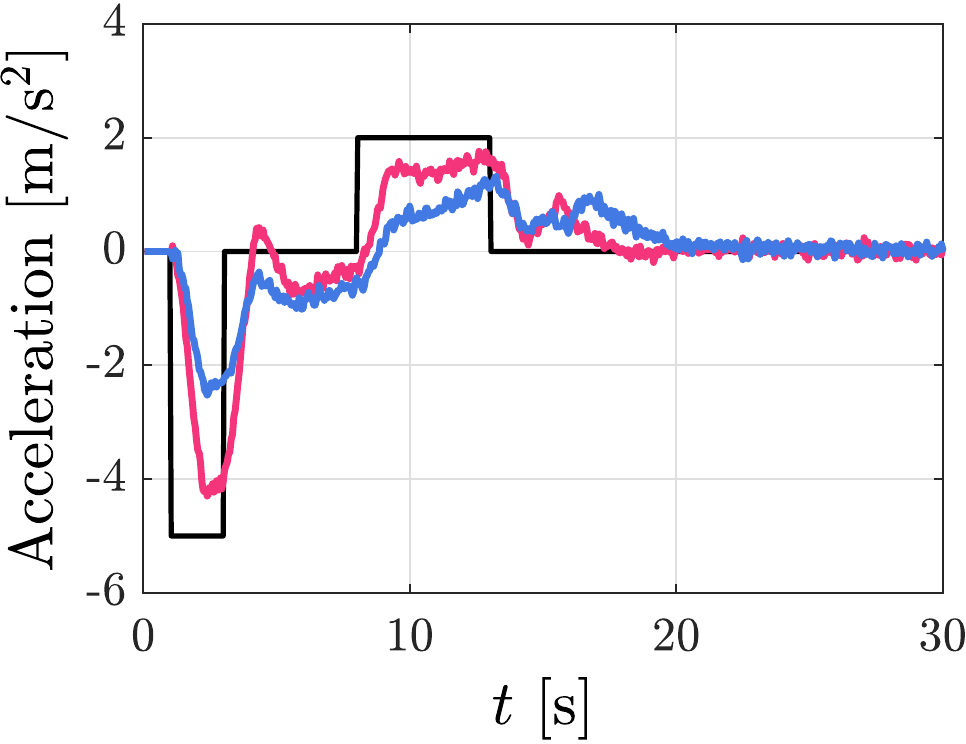}
	\label{Fig:BrakePerturbation_DeePC_Acceleration}}
	\vspace{-1mm}
	\caption{Simulation results in Experiment C, where a sudden brake perturbation is imposed on the head vehicle. (a)(c)(e) demonstrate the velocity, spacing, and acceleration profiles, respectively when all the vehicles are HDVs, while (b)(d)(f) demonstrate the corresponding profiles where there are two CAVs utilizing the  \method{DeeP-LCC} controller. In (c)-(f), the profiles of other HDVs are hided. The color of each profile has the same meaning as that in Fig.~\ref{Fig:SinusoidPerturbation_VelocityProfile}.} 
	\label{Fig:BrakePerturbation_VelocityProfile}
	\vspace{-2mm}
\end{figure}

\subsection{Experiments in Emergence Braking Scenarios}
\label{Sec:Simulation3}

To further validate the safety performance of \method{DeeP-LCC}, we design an emergence braking scenario motivated by Experiment B. As shown by the black profile in Fig.~\ref{Fig:BrakePerturbation_VelocityProfile}, the velocity of the head vehicle is: it maintains the normal velocity at the beginning; then it takes a sudden emergency brake with the maximum deceleration and maintains the low velocity for a while; finally, it accelerates to the original normal velocity and maintains it in the rest time. This is a typical emergency case in real traffic, and it requires the CAVs' control to avoid rear-end collision. { Note that the same pre-collected data set around the equilibrium velocity of $15\,\mathrm{m/s}$ as the previous experiments is utilized in this simulation, and thus as discussed in Section~\ref{Sec:Simulation2} and Appendix~\ref{Appendix:DeePC}, the performance of \method{DeeP-LCC} could still be compromised.}

The results are shown in Fig.~\ref{Fig:BrakePerturbation_VelocityProfile}. 
When all the vehicles are HDVs, they have a large velocity fluctuation in response to the brake perturbation of the head vehicle. By contrast, when two vehicles utilize  \method{DeeP-LCC}, they have a different response pattern from the HDVs: the CAVs decelerate immediately when the head vehicle starts to brake, thus achieving a larger safe distance from the preceding vehicle (see the time period $0-10 \, \mathrm{s}$ in Fig.~\ref{Fig:BrakePerturbation_DeePC_Spacing}); the CAVs also accelerate slowly when the head vehicle begins to return to the original velocity (see the time period $9-12\,\mathrm{s}$ in Fig.~\ref{Fig:BrakePerturbation_DeePC_Acceleration}). In the case of all HDVs, they take a delayed rapid acceleration (see the time period $12-20\,\mathrm{s}$ in Fig.~\ref{Fig:BrakePerturbation_HDVs_Acceleration}), which lead to worse driving comfort and larger fuel consumption. 

{ In addition, for this braking scenario, \method{DeeP-LCC} achieves a comparable performance with respect to MPC which is designed based on prior linearized mixed traffic dynamics. Both strategies save a considerable rate of fuel consumption at a CAV penetration rate of $25\%$ compared with the case of all HDVs (\method{DeeP-LCC}: $24.69\%$, MPC: $25.12\%$). This experiment result further demonstrates the capability of \method{DeeP-LCC}: it allows the CAVs to eliminate velocity overshoot, improve fuel economy, and constrain the spacing within the safe range, whilst requiring no knowledge of HDVs' driving behaviors, contributing to more practical applications than MPC in real-world mixed traffic flow.} 

\balance

\section{Conclusions}
\label{Sec:7}

In this paper, we have presented a novel \method{DeeP-LCC} for CAV control in mixed traffic with multiple HDVs and CAVs coexisting. { Our dynamical modeling and  controllability/observability analysis guarantee the rationality of the data-centric non-parametric representation of mixed traffic behavior in the linearized setup, and further supports the feasibility of applying it to the nonlinear and stochastic mixed traffic system in real scenarios.} In particular, \method{DeeP-LCC} directly relies  on the trajectory data of the HDVs, bypassing a parametric HDV model, to design the CAVs' control input. Multiple numerical experiments confirm that  \method{DeeP-LCC} achieves great improvement in traffic efficiency and fuel economy. 

It is very interesting to adapt our current \method{DeeP-LCC} for time-varying  traffic equilibrium states, in which we need to investigate how to update pre-collected data for constructing Hankel matrices.  Communication delays are another important practical issue to consider in \method{DeeP-LCC}. Existing research have revealed the great potential of standard DeePC in addressing problems with delays~\cite{huang2021decentralized}. {  In addition, the recursive} {  feasibility of  \method{DeeP-LCC} needs further investigation, which guarantees CAV control inputs within safety constraints.} { Finally, computational efficiency of  \method{DeeP-LCC} is worth further investigation for large-scale systems. Similar to distributed MPC~\cite{zheng2016distributed} in CAV control,  distributed versions of \method{DeeP-LCC} will also be extremely interesting. }

\appendices

\section*{Appendices}
\addcontentsline{toc}{section}{Appendices}
\renewcommand{\thesubsection}{\Alph{subsection}}

This appendix provides the proof of Theorem~\ref{Theorem:Controllability}, detailed elaboration of offline data collection, and further discussions on the implementation of  \method{DeeP-LCC}.

\subsection{Proof of Theorem~\ref{Theorem:Controllability}}
\label{Appendix:Controllability}

The following lemma is useful for proving Theorem~\ref{Theorem:Controllability}.

\begin{lemma}[Controllability invariance \rm{\cite{skogestad2007multivariable}}] \label{Lemma:ControllabilityInvariance}
	($A,B$) is controllable if and only if ($A-BK,B$) is controllable for any matrix $K$ with compatible dimensions.
\end{lemma}

Based on Lemma~\ref{Lemma:ControllabilityInvariance}, we transform system ($A,B$) in~\eqref{Eq:LinearSystemModel}
into ($\bar{A},B$) by introducing a virtual input $\bar{u}(t)$, defined as
\begin{equation} \label{Eq:ControlDefinitionTransformation}
    \bar{u}(t) = \begin{bmatrix}u_{i_{1}}(t), \bar{u}_{i_{2}}(t), \ldots, \bar{u}_{i_{m}}(t)\end{bmatrix}^{\top},
\end{equation}
where for $r=2,\ldots,m$, we define
\begin{equation*}
    \bar{u}_{i_{r}}(t) = u_{i_{r}}(t) - \left(\alpha_{1}\tilde{s}_{i_{r}}(t)-\alpha_{2}\tilde{v}_{i_{r}}(t)+\alpha_{3}\tilde{v}_{i_{r}-1}(t)\right).
\end{equation*}
Then, we have
\begin{equation} \label{Eq:ControlTransformation}
    \bar{u}(t) = u(t) - Kx(t),
\end{equation}
where $K = [\mathbb{0}_n,\mathbb{e}_n^{i_2},\ldots,\mathbb{e}_n^{i_m}]^\top \bar{K}$, and $\bar{K}$ is given by
\begin{equation*}
    \bar{K} = \begin{bmatrix} 0 & & &   \\
k_{2,2} & k_{2,1} & &     \\
& \ddots& \ddots&  \\
& &  k_{n,2}&k_{n,1}\\
\end{bmatrix} \in \mathbb{R}^{n\times 2n},
\end{equation*}
with
\begin{equation*}
k_{i,1} = 
\begin{bmatrix}
\alpha_1 & -\alpha_2
\end{bmatrix}, \;
k_{i,2} = 
\begin{bmatrix}
0 & \alpha_3
\end{bmatrix}.
\end{equation*}

According to~\eqref{Eq:ControlTransformation}, we have $A = \bar{A} - BK$. By Lemma~\ref{Lemma:ControllabilityInvariance}, controllability is consistent between ($A,B$) and ($\bar{A},B$). For system ($\bar{A},B$), the physical interpretation of the virtual input $\bar{u}(t)$ in~\eqref{Eq:ControlDefinitionTransformation} is that except the control input of the first CAV, the control input signals of all the other CAVs contain a signal that follows the linearized car-following dynamics of HDVs~\eqref{Eq:LinearHDVModel}. 

Letting $u_{i_{r}}(t)=0$ ($r=2,\ldots,m$), system ($\bar{A},B$) is converted to a mixed traffic system with one single CAV --- only the CAV indexed as $i_1$, \ie, the first CAV in the mixed traffic, has a control input. By Lemmas~\ref{Lemma:CFLCC_Controllability} and~\ref{Lemma:LCC_Controllability}, which state the controllability of the mixed traffic system with one single CAV, system ($\bar{A},B$) thus has the same controllability property. Since the controllability of ($\bar{A},B$) and ($A,B$) are the same, we complete the proof of Theorem~\ref{Theorem:Controllability}.

The proof of Corollary~\ref{Corollary:TransformedSystemControllability} is similar to that of Lemma~\ref{Lemma:LCC_Controllability} when $S=\{1\}$. We refer the interested readers to~\cite{wang2021leading} for details.

\subsection{ Offline Data Collection in \method{DeeP-LCC}}
\label{Appendix:DataCollection}
{

One critical issue in offline data collection of \method{DeeP-LCC} is to guarantee the persistent excitation requirement in Assumption~\ref{Assumption:PersistentExcitation} for the system input, consisting of CAVs' control inputs $u(t)$ and the external input $\epsilon(t)$, \ie, velocity error of the head vehicle. To satisfy this assumption, we present the detailed discussions and the specific implementation method in our simulations below.

\begin{itemize}
    \item For the control input of the CAVs, in practice we need a pre-designed controller (\eg, a car-following model or an ACC-type controller) to control the motion of the CAVs in order to achieve CAV normal driving. Meanwhile, one could add certain i.i.d noise signal into the control model to enrich the control inputs. In our experiments, we utilize the OVM model~\eqref{Eq:OVMmodel} as a pre-designed controller for the CAVs, and the control inputs are designed as
\begin{equation} \label{Eq:ControlInput_DataCollection}
u_i(t)=\alpha\left(v_{\mathrm{des}}\left(s_i(t)\right)-v_i(t)\right)+\beta\dot{s}_i(t)+\delta_u, \quad i \in  S,
\end{equation}
where $\delta_u \in [-1,1]\,\mathrm{m/s^2}$, and the parameters follow the nominal parameter setup in Table~\ref{Tb:HDVParameterSetup}.
    \item For the external input, it is known that in practice, the velocity of the head vehicle is under human control, and it is always oscillating slightly around the human driver's desired velocity. To simulate this scenario, we assume that the external input signal is given by
 \begin{equation}  \label{Eq:ExternalInput_DataCollection}
\epsilon(t)=
    \delta_\epsilon(k)\sim \mathbb{U}[-1,1]\,\mathrm{m/s},
\end{equation}
 where $t=10k+b$ with $k\in\mathbb{N},b\in\{0,1,2,\ldots,9\}$ and $\delta_\epsilon \sim \mathbb{U}[-1,1]\,\mathrm{m/s}$. Recall that in the offline data collection for our experiments, we consider a fixed equilibrium velocity of $15\,\mathrm{m/s}$, \ie, the head vehicle should have a mean velocity of $15\,\mathrm{m/s}$. This design~\eqref{Eq:ExternalInput_DataCollection} means that its velocity changes randomly and slightly around the equilibrium velocity every $10$ time steps ($0.5\,\mathrm{s}$).
\end{itemize}

\begin{figure}[t!]
	\centering
 \subfigure[Control input]
	{\includegraphics[scale=0.42]{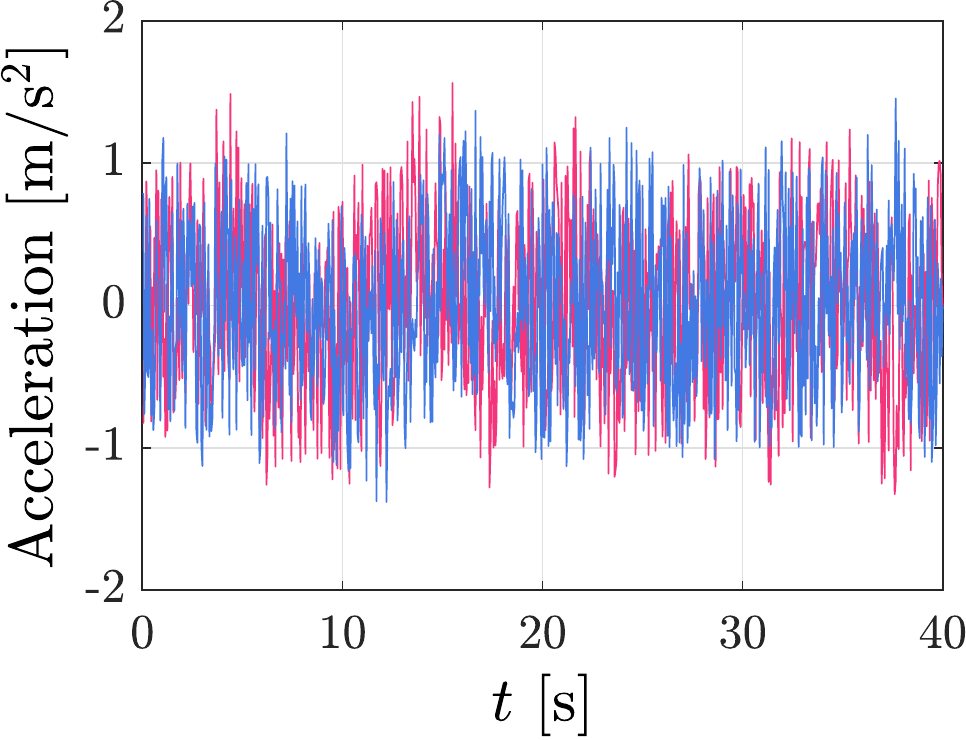}
	\label{Fig:PrecollectedData_ControlInput}}
 \subfigure[External input]
	{\includegraphics[scale=0.42]{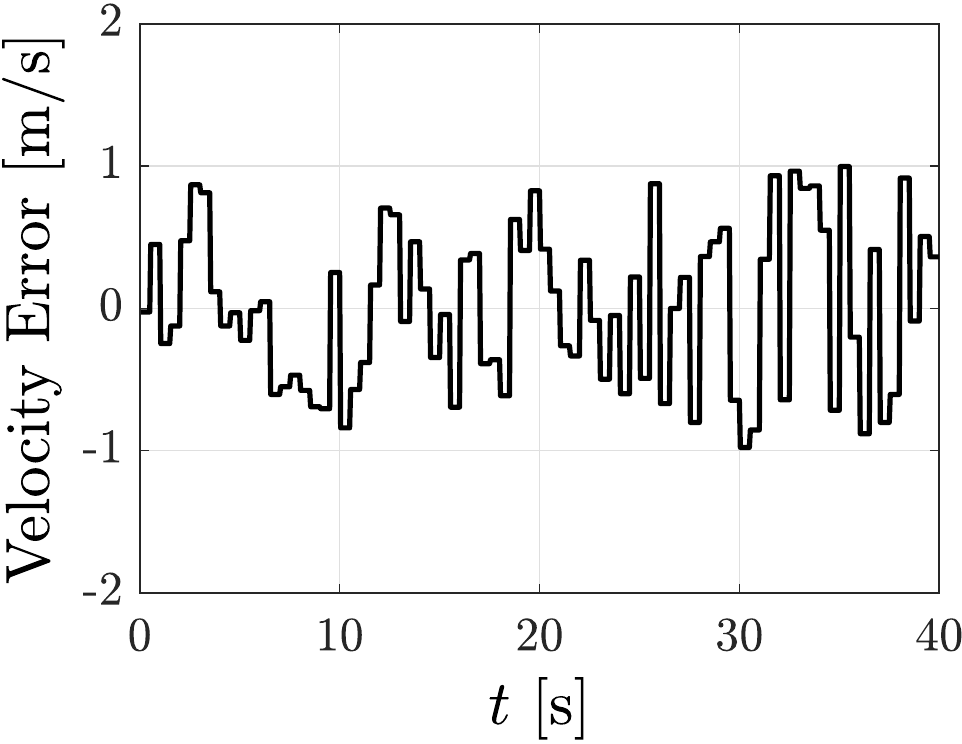}
	\label{Fig:PrecollectedData_ExternalInput}}
	\subfigure[Output (velocity)]
	{\includegraphics[scale=0.42]{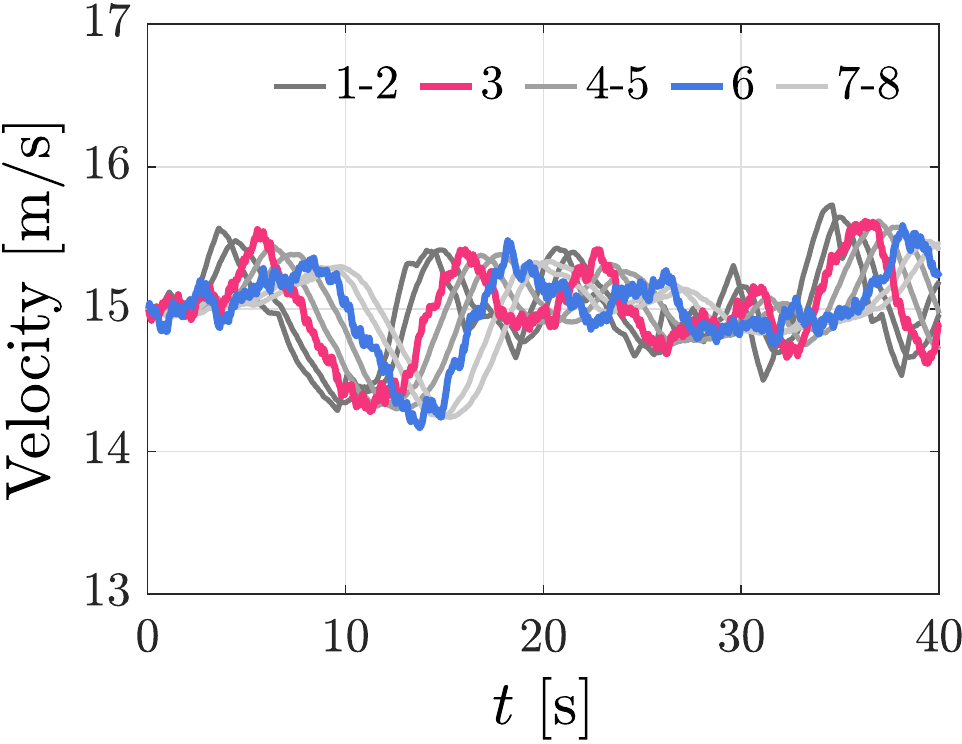}
	\label{Fig:PrecollectedData_OutputVelocity}}
	\subfigure[Output (spacing)]
	{\includegraphics[scale=0.42]{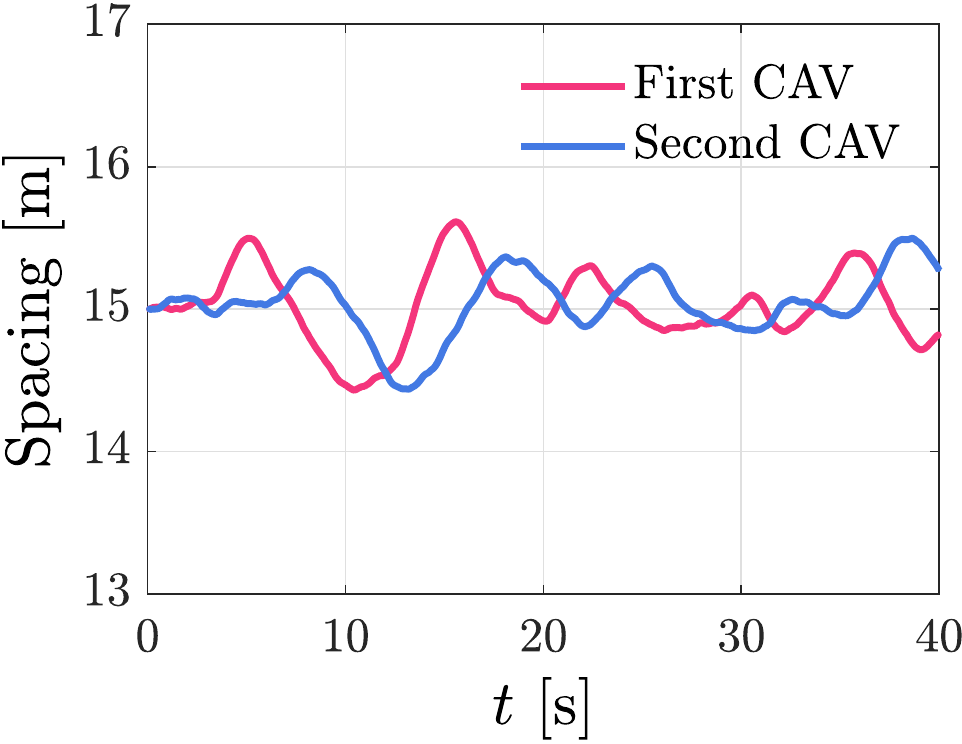}
	\label{Fig:PrecollectedData_OutputSpacing}}	
	\vspace{-1mm}
 \caption{ Illustration of the pre-collected trajectory utilized in Experiments B and C.  (a)(b) demonstrate the system inputs, including the control input, \ie, the acceleration of the CAVs, and the external input, \ie, the velocity error of the head vehicle. (c)(d) demonstrate the measured system output, including the velocity of all the vehicles and the spacing of the CAVs.}
	\label{Fig:PrecollectedData}
	\vspace{-2mm}
\end{figure}

Due to the stochasticity of the input signals~\eqref{Eq:ControlInput_DataCollection} and~\eqref{Eq:ExternalInput_DataCollection}, the persistent excitation requirement can be satisfied when the input length is sufficiently long (since a longer trajectory leads to more columns in the Hankel matrix, making it easier to be full row rank). The theoretical lower bound on the input length $T$ is $(m+1)(T_{\mathrm{ini}}+N+2n)-1$, as revealed in~\eqref{Eq:DataLength}, which is of value $257$ in our experiments in Section~\ref{Sec:6}, and we choose $T=800$ for redundancy considerations.  One could verify the full rank condition by Definition~\ref{Def:PersistentExcitation} before applying the pre-collected data set into controller design. We present one pre-collected trajectory in Fig.~\ref{Fig:PrecollectedData} for illustration, which is also utilized in Experiments B and C in Section~\ref{Sec:6}. 

}

\subsection{Practical Implementation with Time-Varying Equilibrium}
\label{Appendix:DeePC}

In Section~\ref{Sec:Simulation1}, we consider a fixed equilibrium state of $15\,\mathrm{m/s}$ for the simulated traffic flow. The trajectory data is collected around this state and the simulations are also carried out around it. 
{ In Sections~\ref{Sec:Simulation2} and~\ref{Sec:Simulation3}, we have utilized the average velocity of the head vehicle among the past horizon $T_{\mathrm{ini}}$ to estimate the equilibrium velocity. Precisely, 
at time $t$, we have
\begin{equation} \label{Eq:EquilibriumState_Simulation}
    \begin{cases}
    v^*= \displaystyle\frac{1}{T_\mathrm{ini}}\sum_{t-T_\mathrm{ini}}^{t-1}v_0(t),\\
    s^*= \displaystyle\arccos\left(1-2\frac{v^*}{v_{\mathrm{max}}}\right)\cdot \frac{s_\mathrm{go}-s_\mathrm{st}}{\pi} + s_{\mathrm{st}},
    \end{cases}
\end{equation}
where the parameter values follow the nominal setup in Table~\ref{Tb:HDVParameterSetup}. 
This consideration enables the CAV to estimate the real-time equilibrium velocity and meanwhile have a human-like desired spacing policy, according to the OVM model~\eqref{Eq:OVMmodel}. Combining this simple design~\eqref{Eq:EquilibriumState_Simulation} with \method{DeeP-LCC}, our simulation results have revealed the great potential of \method{DeeP-LCC} in improving traffic performance, although~\eqref{Eq:EquilibriumState_Simulation} might lead to mismatched equilibrium states.}

When collecting trajectory data, we still consider the traffic flow around a fixed equilibrium velocity of $15\,\mathrm{m/s}$ to construct the data Hankel matrices. In  \method{DeeP-LCC}, however, we obtain $u_{\mathrm{ini}},y_{\mathrm{ini}}$ by calculating the deviation from the time-varying equilibrium state obtained from~\eqref{Eq:EquilibriumState_Simulation}. { By assuming that the HDVs have a similar behavior around different equilibrium states, one could still apply the fundamental lemma to obtain valid control input.} 

{ This assumption does not always hold in practice, and thus the performance demonstrated in the comprehensive simulation in Section~\ref{Sec:Simulation2} and the braking simulation in Section~\ref{Sec:Simulation3} might not fully reveal the potential of  \method{DeeP-LCC}. Particularly, there could be a mismatch between the current mixed traffic behavior and the predicted behavior generated from the pre-collected data sets by the Willems' fundamental} {  lemma. To address this problem, one approach is to collect trajectory data from multiple equilibrium states, and when implementing  \method{DeeP-LCC}, one can choose appropriate data (\eg, those data around the estimated current equilibrium state) to construct data Hankel matrices and design the control input. Another potential method is to update trajectory data utilized for data Hankel matrices by recording the real-time historical trajectory data in the control procedure. This method is also applicable to the case of time-varying mixed traffic behavior in order to capture the latest dynamics; see, \eg,~\cite{lian2021adaptive} for applications in building control, where the new input/output measurements are appended on the right
side of the Hankel matrices and the old data
on the left side are discarded. Finally, it is also an interesting future direction to investigate the robustness performance of  \method{DeeP-LCC} in mixed traffic in the case of behavior mismatch between data collection and real-time implementation.}



\ifCLASSOPTIONcaptionsoff
  \newpage
\fi



%

\bibliographystyle{IEEEtran}
\bibliography{IEEEabrv,mybibfile}

\end{document}